\documentclass[twocolumn,english,aps,prx,floatfix,amssymb,superscriptaddress,longbibliography]{revtex4}
\usepackage[latin9]{inputenc}
\usepackage{verbatim}
\usepackage{float}
\usepackage{amsmath}
\usepackage{amssymb}
\usepackage{graphicx}
\usepackage{color}
\usepackage{xcolor}
\usepackage{tikz}
\newcommand{\ket}[1]{\ensuremath{\left| #1 \right>}}

\newcommand{\Tr}{\text{Tr}}
\usepackage{bbold}
\DeclareMathOperator{\sgn}{sgn}

\usepackage[bookmarks=false,linkcolor=blue,urlcolor=blue,colorlinks,citecolor=blue]{hyperref}
\makeatletter

\newcommand{\be}{\begin{equation}}
\newcommand{\ee}{\end{equation}}
\newcommand{\bea}{\begin{eqnarray}}
\newcommand{\eea}{\end{eqnarray}}

\begin{document}

\title{Edge Modes in One Dimensional Topological Charge Conserving  Spin-Triplet Superconductors: Exact Results from Bethe Ansatz}
\author{Parameshawar R. Pasnoori}
\affiliation{Department of Physics and Astronomy, Rutgers University, Piscataway, NJ 08854-8019 USA}
\author{Natan Andrei}
\affiliation{Department of Physics and Astronomy, Rutgers University, Piscataway, NJ 08854-8019 USA}
\author{Patrick Azaria}
\affiliation{Laboratoire de Physique Th\'eorique de la Mati\`ere Condens\'ee, Sorbonne Universit\'e and CNRS, 4 Place Jussieu, 75252 Paris, France}

\begin{abstract}
Charge conserving spin singlet and spin triplet superconductors in one dimension are described by the  $U(1)$ symmetric Thirring Hamiltonian.
We solve the model with  open boundary conditions on a finite line segment  by means of the Bethe Ansatz. We show that  the ground state displays a fourfold degeneracy when the bulk is in the spin triplet superconducting phase. This degeneracy  corresponds to the existence  of zero energy boundary bound states localized at the edges which  may be interpreted, in the light of the previous semi-classical analysis due to  Kesselman and Berg \cite{Keselman2015}, as resulting from the existence of fractional spin $\pm 1/4$ localized at the two edges of the system.

\end{abstract}
\maketitle

\section{Introduction}
One of the hallmarks of topological phases  of matter 
is the existence of protected gapless modes localized at their
ends.  This is typically the case of one-dimensional, proximity induced, topological superconductors which host Majorana bound-states at their edges \cite{Alicea2012,Beenakker2013,leijnse2012introduction}.
Since the topological protection of the edge modes  relies
on the presence of a finite energy gap in the bulk, the question
has been raised \cite{Sau2011} whether   purely  one-dimensional superconductors, with the
superconductivity induced  by  intrinsic attractive charge conserving  interactions,
could  support protected localized gapless end modes. Indeed, in these
systems charge conservation   induces strong quantum fluctuations 
leaving  the system gapless with  only quasi-long range superconducting
correlations. Due to the massless charge degrees of freedom, one would 
expect  the edge modes to leak into the bulk and to be only power-law localized
instead of being  exponentially localized.

 However, it has been argued that, provided  there exist 
enough  symmetries, exponentially localized  zero energy  end modes may also be present in gapless systems and  several such phases have been proposed in the literature \cite{Starykh2000, Starykh2007,Ruhman2012,
 Zoller2013, Keselman2015, Diehl2015, chen2017flux, kainaris2017interaction,
 kainaris2017transmission,
  Scaffidi2017,
   Keselman2018}.
Among these, maybe the most  representative example is provided by 
1D spin-triplet topological superconductors (e.g. with dominant triplet superconducting correlations) which are predicted to host  exponentially localized {\it fractional} spin-$\frac{1}{4}$ at both ends of an open chain \cite{Keselman2015} leading to a fourfold ground state degeneracy in the thermodynamic limit.

These predictions are supported by extensive  DMRG calculations \cite{Keselman2015}
in related lattice models, but most of the arguments leading to the existence 
of such  localized fractional  zero energy modes are  based on 
semi-classical or mean-field  arguments. Therefore, we  find it  important  to provide
 an exactly solvable model which displays, over a wide range of coupling constants,  spin
 triplet superconducting correlations in  the bulk and localized zero energy modes  at
 the two edges of an open geometry.   To this end  we shall diagonalize in this work the  
 Hamiltonian  of the  $U(1)$-symmetric Thirring  with Open Boundary Conditions (OBC)
  imposed on the fermions. The Hamiltonian  is given by,
$H= \int_{-L/2}^{L/2} dx\;   {\cal H}$ where 
\bea \label{Hamiltonian}
 {\cal H}&=&  -i v \left(  \psi^{\dagger}_{Ra} \partial_x \psi_{Ra} -  \psi^{\dagger}_{La} \partial_x \psi_{La}\right)
 \\
&+&
   \psi^{\dagger}_{Ra} \psi_{Rb}  [  \,g_{\parallel}\;  \sigma^z_{ab} \sigma^z_{cd} + g_{\perp}\; (\sigma^x_{ab} \sigma^x_{cd}+ \sigma^y_{ab} \sigma^y_{cd})] \psi^{\dagger}_{Lc}\psi_{Ld}.\nonumber 
\eea
In the above equation,  $\sigma^{x,y,z}$ are the Pauli matrices and   the two-components spinor fields  $\psi_{L(R)}(x)$,
which  describe left and right moving  fermions  carrying spin 1/2 with components $a=(\uparrow, \downarrow)$.

The  $U(1)$ Thirring Model, which  is an anisotropic $XXZ$-type  deformation
of the $SU(2)$ invariant Thirring model (or the Chiral invariant  two-flavors Gross-Neveu model),
describes both singlet and spin triplet  1D charge conserving superconductors
as well as  the quantum phase transition between them,  as function of the couplings $(g_{\parallel}, g_{\perp})$.

The model  has  has been shown to be integrable  with Periodic Boundary Conditions (PBC)  \cite{Andrei, duty, Japaridze}. However we are not aware of a solution of the model   on a finite line segment with OBC,
\be
\psi_{La}(\pm L/2) + \psi_{Ra}(\pm L/2)= 0, \; a=(\uparrow, \downarrow).
\label{OBC}
\ee
It is only, to our knowledge,  in the  $SU(2)$ invariant case (i.e: $g_{\parallel}=g_{\perp}$) that an exact solution has been obtained recently  on the system with one open edge with a Kondo impurity  coupled to it  \cite{PRA1}.
As we shall demonstrate in  section (\ref{sec:betheAnsatz}), the   model   is  integrable for arbitrary couplings  when the OBC (\ref{OBC})  are imposed on the fermions and it remains integrable also in the presence of more general  boundary conditions which are  {\it asymmetric}
with respect to the left and the right edges.

  Solving the model we find it possesses both  topologically trivial and non trivial phases
  corresponding to spin singlet and spin triplet superconducting correlations respectively.
   While the topologically trivial SSS phase phase is unique and nondegenerate a more interesting situation arises in the topological STS phase. The exact solution shows that for an infinitesimal asymmetric OBC  a four degenerate ground state structure emerges resulting from the existence  of two zero energy boundary bound states localized at the two ends of the system.
Two of the ground states have a total z-component spin $S^z=0$ and  fermion parity ${\cal P}=+1$ while
the other two have $S^z=\pm 1/2$ and  fermion parity ${\cal P}=-1$. This  four-fold degeneracy  
can be consistently interpreted in the light of the semi-classical analysis \cite{Keselman2015}
as {\it fractional} spin-$\frac{1}{4}$ boundary states.  We remark however that when the asymmetry between the left and right edges is removed we explicitly find only  a  {\it threefold} degenerate ground state in the thermodynamic limit. We shall argue that the fourth state can be obtained by acting with 
a symmetry operator, analogous to the spin lowering operator that needs to be applied to the highest weight spin state provided by the solution to the Bethe Ansatz equations of SU(2) symmetric models in order to complete the multiplets. In the topological phase such an operator would be given by either  one of the two zero energy Majorana modes,
localized at the two edges of the system,
characterizing the topological degeneracy in
a given fermionic parity sector.
Overall, our exact results indicate that quantum fluctuations do not spoil 
the topological nature of the spin triplet phase found in the semi-classical limit.

\smallskip

The paper is organized as follows.  We begin in Section (\ref{sec:bulk}) by reviewing the bulk properties of (\ref{Hamiltonian}) using both fermionic and bosonic languages. In the section (\ref{sec:boundary}) we 
elaborate on the semi-classical arguments given by Keselman and Berg \cite{Keselman2015} leading to existence of localized  {\it fractional} spin-$\frac{1}{4}$ boundary states in the spin triplet superconducting phase. In section  (\ref{sec:betheAnsatz}) we solve the model using   Bethe Ansatz  in the scaling limit where universal answers can be obtained  for the ground state as well as for boundary 
 excitations.   We shall also consider the effect of integrable asymmetric  boundary conditions between  left and right edges.  We finally discuss our results and open questions in the the section (\ref{sec:discussion})

\section{Bulk Properties of the $U(1)$-symmetric Thirring model}
\label{sec:bulk}
In this section we shall present, to be self-consistent, some of the known results regarding the  model. They follow from the exact solution given in Ref. \cite{ Japaridze}
for PBC and bosonization. Most of these results hold when OBC are considered
as far as bulk properties are concerned.

\subsection{Symmetry Properties}

We start by briefly discussing the symmetry properties of the model. The Hamiltonian (\ref{Hamiltonian}) displays,  for generic couplings $(g_{\parallel},  g_{\perp})$, a  $U(1)_c \otimes U(1)_s$ symmetry corresponding to the changes, $\psi_{L(R)} \rightarrow e^{i \alpha_c} \; \psi_{L(R)}$ and $\psi_{L(R)} \rightarrow e^{i \alpha_s \sigma^z/2} \; \psi_{L(R)}$.
As a consequence the total spin $S^z$ and the total number of fermions $N$,
\bea
S^z&=&  \frac{1}{2}\int dx\;  \left( \psi^{\dagger}_{L}\sigma^z\psi_{L} +  \psi^{\dagger}_{R}\sigma^z\psi_{R} \right), \\
\label{spin}
N &=&  \int dx\;  \left( \psi^{\dagger}_{L}\psi_{L} +  \psi^{\dagger}_{R}\psi_{R}\right),
\label{number}
\eea
are conserved quantum numbers. The model  displays also a number of discrete symmetries.
On top of the chiral symmetry, i.e:  $\psi_{L(R)} \rightarrow \psi_{R(L)}$,  (\ref{Hamiltonian})
is  time-reversal (TR) symmetric,
$
 {\cal T}\psi_{L(R)}= i\sigma^y \psi_{R(L)}$ ($ {\cal T}^2=-1
$),  and is invariant under space parity $x \rightarrow -x$. The $U(1)$-Thirring model
is  also invariant upon reversing  the spins of all the fermions, i.e:  $ \Psi_{L(R),\uparrow} \leftrightarrow \Psi_{L(R),\downarrow}$.  The latter symmetry has a  $\mathbb{Z}_2=\{1, \tau \}$ group structure
with
\be
\tau \psi_{L(R)} =\sigma^x \; \psi_{L(R)}, \; \tau^2=1,
\label{z2}
\ee
and, similarly to time reversal ${\cal T}$, it reverses the sign of the total spin $S^z$, $S^z  \rightarrow -S^z$.
Finally,  on the  line  $g_{\parallel}=  g_{\perp}$,  the $U(1)_s$ symmetry in the spin sector  is  enlarged to $SU(2)$. 
On this line, the $U(1)$-Thirring model  (\ref{Hamiltonian}) is invariant under the shift, 
$
\psi_{L(R)} \rightarrow e^{i \vec \alpha \cdot \vec \sigma/2} \; \psi_{L(R)}
$,
and is nothing but the    Gross-Neveu (GN) model. 

\paragraph{Duality Symmetry.} On top of the above symmetries, the model displays also a 
{\it duality} symmetry $\Omega$ \cite{BoulatDuality} which acts asymmetrically  on the left and right fermions
\bea
\ \widehat  \psi&=& \Omega\psi \nonumber \\
 \widehat \psi_{L}= \psi_{L}&,& \; \widehat \psi_{R} = i  \sigma^z  \psi_{R}.
\label{duality}
\eea
The duality $\Omega$  relates exactly, and at all scales, different models with  opposite couplings  $g_{\perp}$ and  $-g_{\perp}$:
\be
{\cal H}(\psi, g_{\parallel},  g_{\perp}) = {\cal H}(\widehat \psi, g_{\parallel},  -g_{\perp}).
\label{dualHamiltonian}
\ee
In particular  it relates  the correlation functions between 
any set of operators ${\cal O}_j (\psi)$ and their duals $\widehat { \cal O}_j \equiv {\cal O}_j(\widehat  \psi)$ in the two ground states of ${\cal H}(\psi, g_{\parallel},  \pm g_{\perp}))$, i.e:
$ \langle {\cal O}_1...{\cal O}_n \rangle_{g_{\perp}} = \langle \widehat { \cal O}_1...\widehat{\cal O}_n \rangle_{-g_{\perp}} $. This property allows in principle to deduce the properties 
of dual models with couplings with say $(g_{\parallel}, - g_{\perp} <0 )$ from those
of models with couplings $(g_{\parallel},  g_{\perp}  >0 )$.
It is a symmetry of the phase diagram.
Of  particular interest  is the model described by (\ref{Hamiltonian}) on the line $g_{\parallel}= - g_{\perp}$
which is dual to the  $SU(2)$  Gross-Neveu model ${\cal H}_{\rm GN}(\psi_a, g,  g)$ and displays a dual
$\widehat{SU(2)}$ non-local symmetry
\be
\psi_{L} \rightarrow e^{i \vec \alpha \cdot \vec \sigma/2} \; \psi_{L},\; \psi_{R} \rightarrow (\sigma^z e^{i \vec \alpha \cdot \vec \sigma/2} \sigma^z)  \psi_{R}.
\label{su2dual}
\ee
We shall refer to the model described by Hamiltonian
${\cal H}_{\rm \widehat{GN}}(\psi, g,  -g)$  as the  dual  $\widehat{SU(2)}$ Gross-Neveu 
model that we shall denote $\widehat{GN}$ in the following.

\subsection{ Bosonization}

The Hamiltonian  (\ref{Hamiltonian}) may also be expressed in terms of bosonic fields using the equivalence between the $U(1)$-Thirring model and the sine-Gordon (SG) model. The latter correspondance  is valid  in the long distance and low energy limit \cite{Coleman1975, Japaridze} and is achieved  using the bosonization of the Fermi field
\bea
 \psi_{L(R)a}(x) &=& \frac{\kappa_a}{\sqrt{2\pi a_0}} \exp{[ -i \sqrt{ \pi}(\theta_{a}(x) \pm \phi_{a}(x) )]}, 
 \label{bosonization}
 \eea
where $a=(\uparrow, \downarrow)$,  $a_0$ is a short distance cutoff and $[\phi_a(x), \theta_b(y)]= -i \delta_{ab} H(x-y)$, $H(u)$ being the Heaviside step function. The operators  $\kappa_{a=(\uparrow, \downarrow)}$ are anticommuting Klein factor satisfying $\{\kappa_a, \kappa_b\}=2\delta_{ab}$ which insure that fermions of different spins anticommute. The long distance and low energy limit of  (\ref{Hamiltonian}) is described  by two scalar bosons fields, a massless charge field $\Phi_c$ and a SG spin field $\Phi_s$ as well  as their duals $\Theta_c$ and $\Theta_s$,
\bea
\Phi_{c}= (\phi_\uparrow + \phi_\downarrow)/\sqrt{2}&,& \; \Theta_{c}= (\theta_\uparrow  + \theta_\downarrow)/\sqrt{2} \label{chargefield} \\
\Phi_{s}=\frac{2\sqrt{\pi}}{\beta} (\phi_\uparrow - \phi_\downarrow) &,& \; \Theta_{s}=\frac{\beta}{4 \sqrt{\pi}} (\theta_\uparrow  + \theta_\downarrow),
\label{spinfield}
\eea
in terms of which the  total number of fermions $N$ and the total spin $S^z$ can be expressed  as,
\be
N=\sqrt{\frac{2}{\pi}} \int dx\; \partial_x\Phi_c, \; S^z = \frac{\beta}{4\pi}  \int dx\; \partial_x\Phi_s.
\label{chargespinboso}
\ee 

 The Hamiltonian (\ref{Hamiltonian}) written in terms of these low energy fields,
decomposes as ${\cal H}= {\cal H}_{LL} + {\cal H}_{SG}$
where 
\bea
{\cal H}_{LL} &=& \frac{u_c}{2}[\frac{1}{K_c} (\partial_x\Phi_c)^2 + 
 K_c (\partial_x\Theta_c)^2], 
 \label{luttinger}
 \eea 
 a Luttinger liquid Hamiltonian which describes the gapless charge degrees of freedom and
\bea
{\cal H}_{SG} = \frac{u_s}{2}[ (\partial_x\Phi_s)^2 + 
 (\partial_x\Theta_s)^2] 
 - \chi  \frac{m_0^2}{\beta^2} \cos{(\beta \Phi_s)}, 
 \label{sineGordon}
\eea
is the  sine-Gordon (SG) Hamiltonian which describes the spin
degrees of freedom. Although in the integrable $U(1)$ Thirring model the charge Luttinger parameters  $u_c=v$ and $K_c=1$, for sake of generality, we shall keep them as generic parameters in the following. In the spin sector $\chi = \sgn{(g_{\perp})}$ and 
the couplings   $u_s$, $m_0^2$ and $\beta$ of the SG model (\ref{sineGordon}) are related  to the  couplings $(g_{\parallel},  g_{\perp})$ of the $U(1)$ Thirring model (\ref{Hamiltonian}) in a  non-universal way except in  
the weak coupling limit $(|g_{\parallel}|,  |g_{\perp}|) \ll 1$ where $m_0^2/\beta^2  \approx g_{\perp}u_s/(\pi a_0)^2$  and $\beta^2/8\pi \approx 1-g_{\parallel}/(\pi v)$.

Finally, to be complete,  let us quote how the discrete symmetries of the model act on the boson fields. 
The $\mathbb{Z}_2$ symmetry generator $\tau$ in (\ref{z2}) acts on the spin boson fields as 
\be
\tau \Phi_s = - \Phi_s, \, \tau \Theta_s= -\Theta_s,
\label{z2boson}
\ee
while the duality symmetry $\Omega$ (\ref{duality}) on the fermion fields translate as
\be
\Omega \Phi_s = \Phi_s + \frac{\pi}{\beta}, \; \Omega \Theta_s = \Theta_s - \frac{\pi}{\beta}.
\label{dualityboson}
\ee

\smallskip

\subsection{Phase Diagram}
\label{sec:phasediagram}
\begin{center}
\begin{figure}[!h]
\includegraphics[width=0.6\columnwidth]{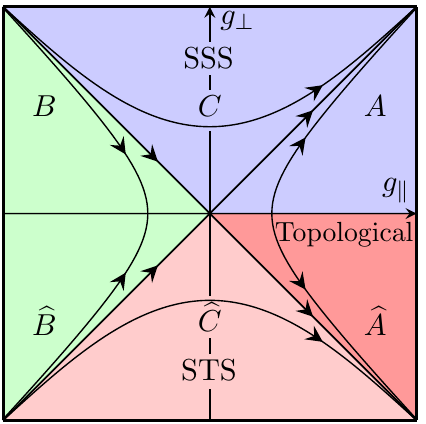}
\caption{Weak-coupling  phase diagram of the $U(1)$ Thirring  model. In green is the the Luther-Emery phase
which includes the regions B and $\widehat{B}$. In blue is the  Spin Singlet Superconducting (SSS) phase. It includes the two regions  A and C as well as the $SU(2)$ invariant GN line at $g_{\parallel}=g_{\perp}$. In red is the 
Spin Triplet Superconducting (STS) phase.
It includes the two regions $\widehat{A}$ (dark red) and $\widehat{C}$ (light red) and the $\widehat{GN}$ line at $g_{\parallel}=-g_{\perp}$. The semi-classical
regime corresponding to large $g_{\parallel} \gg 1$ is not displayed in the figure. In a system with OBC we show using Bethe Ansatz 
that, in the region A and on the GN line of the SSS phase,  the ground state is non degenerate. In contrast, in the region $\widehat{A}$ and on the $\widehat{GN}$ line of the STS phase (dark red) the ground state is fourfold degenerate.
This degeneracy results from zero energy  boundary bound states localized at the edges. We interpret them  in the light of  the semi-classical analysis of Kesselman and Berg \cite{Keselman2015} as the result of localized spin $\pm 1/4$ at the two edges of the system. We  hence give support that the topological phase found in the strong anisotropic regime in \cite{Keselman2015} survives quantum fluctuations down to the weak coupling regime. 
In the regions C and $\widehat{C}$ we are unable to conclude from the Bethe Ansatz analysis about the ground state degeneracy in the universal regime.
\label{fig:PD}}
\end{figure}
\end{center}

The phase diagram of the $U(1)$ Thirring model depicted in  Fig.\ref{fig:PD},  is well known and was obtained in Ref. \cite{ Japaridze} in the case of  periodic boundary conditions. The plane $(g_{\parallel},  g_{\perp})$ is divided into six regions. For  $g_{\perp} > 0$ one distinguishes between three regions: A for $g_{\parallel} > g_{\perp} > 0$, B for $g_{\parallel} < - g_{\perp} < 0$, and C for $|g_{\parallel}| > g_{\perp} $.  For $g_{\perp} < 0$
the other three  regions are:   ${\widehat A}$ for $g_{\parallel} > - g_{\perp} > 0$,  ${\widehat B}$ for $g_{\parallel} < g_{\perp} < 0$, and  ${\widehat C}$  for $|g_{\parallel}| <  - g_{\perp} $. They
are the dual to the  $g_{\perp} > 0$ regions  in the sense of (\ref{duality}).  On top of these six 
regions  are the two invariant $GN$ and $\widehat{GN}$ lines obtained for  $g_{\parallel}=  g_{\perp}$ and $g_{\parallel}=  -g_{\perp}$. On these lines, the model displays  enlarged  $SU(2)$ and $\widehat{SU(2)}$ symmetries.
 
In the regions B and ${\widehat B}$  the four fermion term, proportional to $g_{\perp}$ in (\ref{Hamiltonian}), is irrelevant and both  spin and charge sectors remain massless.
The low energy sector of the theory is described by two Luttinger liquids, one in each sector. This is the Luther-Emery phase \cite{ LutherEmery}. The same conclusion holds for the 
$GN$ and $\widehat{GN}$ lines when $g_{\parallel} <  0$.

In  contrast,  in the remaining regions, A (${\widehat A}$),   C (${\widehat C}$),  and on the  $GN$ ($\widehat{GN}$) lines for $g_{\parallel} >0$, the four fermion  term is relevant:   there is a dynamical mass generation $m$,  corresponding to the opening of a spin  gap, while  a massless charge excitation decouples from the spectrum. When $g_{\perp} > 0$, the  (A, C) regions and the  $GN$ line define the same phase.  When $g_{\perp} < 0$ the two regions  ${\widehat A}$,  ${\widehat C}$ and the $\widehat{GN}$ line define a different, dual, spin gapped phase. The reason why we distinguish between two regions in a single phase is  that, while the theory is asymptotically free in the ultraviolet in the regions A(${\widehat A}$) as well as on the $GN$ and $\widehat{GN}$ lines, in regions C(${\widehat C}$), it is non trivial in both the infrared and the ultraviolet regimes. In all these regions the  (massive) spectrum and  the  Bethe  equations are different but the ground state properties do not qualitatively change with in each of the two spin gapped phases.   In particular we stress that   there are no phase transitions on either  the $GN$ and $\widehat{GN}$ lines between the regions A and C or ${\widehat A}$ and ${\widehat C}$. The only phase transition  between the two spin gapped phases is between regions A and ${\widehat A}$ one the line $g_{\perp}=0$ where the $U(1)$ Thirring model is described by two charge and spin Luttinger liquids. We refer the reader interested for more details in these topics to Ref.\cite{ Japaridze}. 

\smallskip

\subsubsection{Ground State Instabilities}

As well known in one dimension, due the massless  charge degree of freedom, there can be no true long range order of any local, superconducting and/or charge density wave types. The different phases are instead characterized by both
the ground state degeneracy and the dominant instabilities (i.e. the power-law asymptotics of their correlation functions) they support. In all the phases, the ground state is not degenerate for PBC and we distinguish between two spin gapped phases.

{\it -The SSS  phase.} This is a   Spin Singlet Superconducting  phase which  is stabilized when $g_{\perp} > 0$ and corresponds to   the regions A, C  and the $GN$ line. In this phase a non-zero mass gap $m$ develops  and the system 
displays both  Spin Singlet Superconducting   and  Charge Density Wave (CDW) instabilities  with order parameters,
\bea
{\cal O}_{\rm SSS} = \psi^{\dagger}_{L} \sigma^y  \psi^{\dagger}_{R},\; 
{\cal O}_{\rm CDW} = \psi^{\dagger}_{L} \;   \psi_{R}.
\label{SSSCDW}
\eea
which, using bosonization, can be shown to display  quasi-long-range order  \bea
\langle {\cal O}^{\dagger}_{\rm SSS}(x) {\cal O}_{\rm SSS}(0) \rangle &\propto &|x|^{-1/K_c}, \\\label{corrsss}
\langle {\cal O}^{\dagger}_{\rm CDW}(x) {\cal O}_{\rm CDW}(0) \rangle  &\propto &|x|^{-K_c},
\label{corrcdw}
\eea
where $K_c$ is the charge Luttinger parameter.

{\it -The STS  phase.} This is a Spin Triplet Superconducting phase which is obtained when  $g_{\perp} <  0$ in  the regions  $\widehat{ A}$,  ${\widehat C}$ and $\widehat{GN}$. In this phase  both SSS and CDW  correlation functions are short range and the instabilities are of the  Spin Triplet Superconducting (STS) and the Spin Density Wave (SDW)  types with order parameters,
\bea
{\cal O}_{\rm STS} &=& \psi^{\dagger}_{L} \sigma^x  \psi^{\dagger}_{R}, \; 
{\cal O}_{\rm SDW} = i \psi^{\dagger}_{L}  \sigma^z  \psi_{R},
\label{STSSDW}
\eea
and asymptotics
\bea
\langle {\cal O}^{\dagger}_{\rm STS}(x) {\cal O}_{\rm STS}(0) \rangle &\propto &|x|^{-1/K_c}, \\\label{corrsts}
\langle {\cal O}^{\dagger}_{\rm SDW}(x) {\cal O}_{\rm SDW}(0) \rangle  &\propto &|x|^{-K_c}.
\label{corrsdw}
\eea
The two types of instabilities in (\ref{SSSCDW}) and (\ref{STSSDW}) are mutually non-local and are actually dual
to each other in the sense of (\ref{duality}), i.e: ${\cal O}_{\rm STS}= \widehat{{\cal O}_{\rm SSS}}$ and ${\cal O}_{\rm SDW}= \widehat{{\cal O}_{\rm CDW}}$. They  only coexist on the line of fixed points  $g_{\perp} = 0$  where  the spin gap closes.  The two SSS and STS phases define therefore two  different phases separated by the  quantum phase transition line  ($g_{\perp} = 0, g_{\parallel}\ge 0$).  Notice that in a general 1D electron gas \cite{Giammarchi}  the superconducting and density waves instabilities have different power-law asymptotics, i.e: $\propto |x|^{-1/K_c}$ and $\propto |x|^{-K_c}$ respectively.  Which instability dominates depends on $K_c$ and it is custom to label the  phases by their dominant instability: either CDW or SDW  when $K_c < 1$ and either SSS or STS phase when $K_c > 1$. In the  present integrable model, where $K_c=1$,  both types of instabilities are equally dominant and we choose to label the two  phases by the type of superconducting instability they support, i.e:   SSS  when $g_{\perp} > 0$ and STS when  $g_{\perp} < 0$.

\section{Open Boundaries: semi-Classical approximation}
\label{sec:boundary}

We now consider the effect of the OBC (\ref{OBC}) on the fermions. Before going into the detailed analysis of
our exact solution, we discuss  the  model in the strongly anisotropic  limit: $g_{\parallel} \gg 1, |g_{\perp}| \ll 1$ (called quasi-classical regime in \cite{Japaridze}) where, as far as bulk properties are concerned, the $U(1)$ Thirring model can be regarded  as a regularized  integrable version of the SG model (\ref{sineGordon})  with $\beta^2/4\pi < 1$ and $m_0^2$ small. In the quantum regime, i.e: when 
$1 < \beta^2/4\pi < 2$, this ceases to be true and the equivalence between the $U(1)$ Thirring model
and the SG model is only valid in the asymptotic low-energy/long-distance limit.

The OBC on the fermions (\ref{OBC}) translate to the charge bosonic field (\ref{luttinger}) and  the spin bosonic field of the SG model  (\ref{sineGordon}) as
\bea
 \Phi_c(-L/2)&=&0, \; \Phi_c(+L/2) =\sqrt{\frac{\pi}{2}} N, \\
\Phi_s(-L/2)&=&0, \; \;  \Phi_s(+L/2) = \frac{4\pi}{\beta} S^z.
\label{OBCspinboson} 
\eea
Therefore,  since the total number of particle of each spin  component $N_{(\uparrow, \downarrow)}$  enclosed in a finite chain is an integer,  $N \pm 2 S^z$ has to be an  even integer. This implies that  the total particle number sector  $N$  of the Luttinger liquid Hamiltonian (\ref{luttinger}) and the total  $S^z$ sector of the SG model  (\ref{sineGordon})  
 are not independent but constrained by
\be
N \; {\rm even} \Leftrightarrow S^z \in \mathbb{Z}, \; N \; {\rm odd} \Leftrightarrow S^z \in \mathbb{Z} +\frac{1}{2}.
\label{Nevenodd}
\ee
Besides the latter constraint, the physics of the  $U(1)$ Thirring model in the presence of OBC in the semi-classical regime boils down to that of the SG model (\ref{sineGordon}) with the  OBC (\ref{OBCspinboson}). The effect of Dirichlet
boundary conditions on the spectrum of the SG model has been extensively studied
\cite{gosh, gosz, skorik}. It appears that, in presence  of the OBC (\ref{OBCspinboson}),  boundary bound states
are likely to be stabilized in an extended part of the phase diagram which corresponds to a region 
which is deep in the STS phase. As we shall see, these boundary bound states emerge straightforwardly 
in the semi-classical analysis and  are responsible, as first stressed out in Ref.  \cite{Keselman2015},  of the topological nature of the STS phase in the limit of large $g_{\parallel} \gg 1, |g_{\perp}| \ll 1$.

\subsection{Semi-Classical Analysis}

The semi-classical limit of the SG  model   corresponds to  the limit $\beta \rightarrow 0$ (keeping $m_0^2$ small fixed)
and  to large $g_{\parallel} \gg 1$. It is well understood as far as PBC are concerned \cite{Rajaraman}. For OBC, as we shall see, the ground state degeneracy dramatically changes when going from the SSS phase ($g_{\perp}> 0$) to the STS phase $g_{\perp}> 0$. In the limit $\beta \rightarrow 0$, as  argued by Keselman and Berg \cite{Keselman2015},  the system in the STS phase hosts  symmetry protected   edge states with fractional spin $S^z = \pm 1/4$, which are exponentially  localized at each edge. These edge states  become degenerate in the thermodynamical limit leading  to a fourfold ground state degeneracy. In the following we shall follow the lines of arguments presented in Ref. \cite{Keselman2015} and elaborate on the nature of the edge states Hilbert space.


\subsubsection{Classical Edge Kinks}

In the  limit, $\beta \rightarrow 0$,   the boson spin  field is  locked to  the minima $\Phi_{s,n}$ of the cosine term in (\ref{sineGordon}) which   depend on the sign of  $g_{\perp}$
\bea
\Phi_{s,n}&=&n \frac{2\pi}{\beta}, \; \; \; \; \; \; \; \;  \; \; \;  \chi > 0, \\
\label{SG+groundstates}
 \Phi_{s,n}&=&(n + \frac{1}{2}) \frac{2\pi}{\beta}, \; \chi < 0,
 \label{SG-groundstates}
\eea
 where we recall that $\chi=\sgn(g_{\perp})$ and  $n\in \mathbb{Z}$. Clearly in the SSS phase ($\chi >0$) given the OBC (\ref{OBCspinboson}),  there is a unique minimum at $\Phi_{s,n}=0$ which corresponds to a total spin $S^z=0$.
 In the STS phase ($\chi < 0$)  the situation radically changes
as none of the minima in (\ref{SG-groundstates}) match with the OBC (\ref{OBCspinboson}).
The lowest energy states in this case  consist of classical kinks configurations of the spin field $\Phi_s(x)$  which interpolate between  $\Phi_s(-L/2)=0$ and $\Phi_s(+L/2)= \frac{4\pi}{\beta} S^z$ and   match  in the bulk (i.e. when $-L/2 \ll x \ll +L/2$) with one of the   classical ground states (\ref{SG-groundstates}).   Due to the spin gap $m$ in the bulk, these kinks   are exponentially localized near both left and right  edges  
 at  $x=-L/2$ and $x=+L/2$ respectively and, to the  exponential accuracy in the system size 
(i.e: to  ${\cal O}(e^{-mL})$), each edge can be treated separately.
 Hence, for large system sizes, i.e: when $mL \gg 1$, 
 the kinks can be seen as the sum of left and right kinks which interpolate
 between $\Phi_s(-L/2)=0$ and $\Phi_{s,n}=(n + \frac{1}{2}) \frac{2\pi}{\beta}$ in bulk and 
 between $\Phi_{s,n}=(n + \frac{1}{2}) \frac{2\pi}{\beta}$ in bulk and 
  $\Phi_s(+L/2)= \frac{4\pi}{\beta} S^z$.

\begin{center}
\begin{figure}[!h]
\includegraphics[width=0.6\columnwidth]{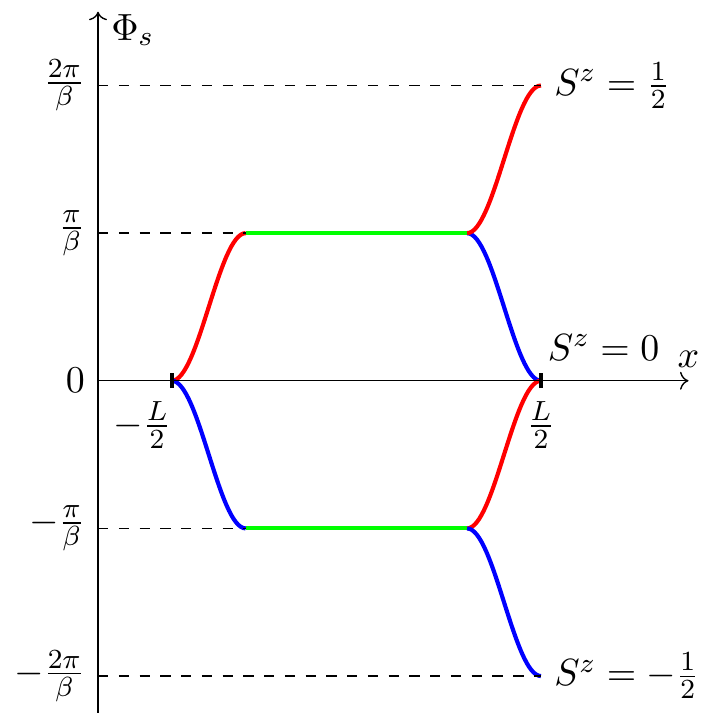}
\caption{Classical kink configurations in the  STS phase in the presence OBC as taken from Ref.\cite{Keselman2015}. The red (blue) lines correspond to edge  kinks with an accumulation of spin $\Delta S^z=\frac{1}{4}$($-\frac{1}{4}$). The two green horizontal lines correspond to the constant values that  the spin field $\Phi_s$ takes in the two bulk ground states $\Phi_s=\pm \frac{\pi}{\beta}$. \label{fig:kinks}}
\end{figure}
\end{center}

Consider first the left edge  at $x=-L/2$. As depicted in the Fig.(\ref{fig:kinks}), there
the lowest energy states consist of classical kinks  interpolating between    $\Phi_s(-L/2)=0$
and one of the two classical ground states   $\Phi_{s}=\pm \pi/\beta$ in the bulk. 
Due to (\ref{chargespinboso}) these two left  kinks  correspond to an accumulation of a {\it fractional} spin at the left edge,
\be
S_{\cal L}^z=  \frac{\beta}{4\pi}  \int_{-L/2}^{y} dx\; \partial_x\Phi_s =\pm 1/4,
\label{SzL}
\ee
 where $y$ is some point deep in the bulk. These two left kinks
cost a finite energy but have the same energy owing to the $\mathbb{Z}_2$ symmetry (\ref{z2boson}) which
exchanges the two ground states in the bulk, i.e: $ \pi/\beta \leftrightarrow - \pi/\beta$.
A similar analysis can be done  at the right edge at $x=L/2$. The situation there  
depends on the total spin $S^z$ enclosed in the system. When $S^z=0$ the lowest energy states   consist of right  kinks   interpolating between      $\Phi_s=\pm \pi/\beta$ in the bulk and  $\Phi_s(+L/2)= 0$. They correspond to an accumulation of a fractional spin, 
 \be
S_{\cal R}^z=  \frac{\beta}{4\pi}  \int_{y}^{L/2} dx\; \partial_x\Phi_s =\mp 1/4,
\label{SzR}
\ee
at the right edge. When $S^z=\pm 1/2$ the right kinks  interpolate
between $\Phi_s=\pm \pi/\beta$ in the bulk  and $\Phi_s(+L/2)= \pm 2\pi/\beta$ at the edge. They also accumulate a fractional spin a  the right edge $S_{\cal R}^z=\pm 1/4$. Since kinks depends only  the accumulation of  spin they carry, there are only two independent right classical  kinks  with spin $\pm 1/4$.
 Due tho the $\mathbb{Z}_2$ symmetry (\ref{z2boson}) these right kinks have the same energy.
 All together there are four  classical kinks states, two at the  left and two at the right edges, each carrying 
fractional spins $\pm 1/4$ and having, to ${\cal O}(e^{-mL})$ accuracy, the same classical energy thanks to the $\mathbb{Z}_2$  symmetry (\ref{z2boson}).

\subsubsection{Quantum Edge Kinks.}

 To  promote  the  above  the left and right classical kinks to quantum states one needs to  {\it  assume}  
 the existence of quantum kinks at each edge
\be
|\pm \frac{1}{4} \rangle_{\cal L}, \; |\pm \frac{1}{4} \rangle_{\cal R},
\label{edgekinks}
\ee
 labelled by their local fractional  spins.
Concurrently this implies the
existence of local  quantum spin operators $S_{\cal L}^z$ and $S_{\cal R}^z$, $[S_{\cal L}^z, S_{\cal R}^z]=0$, which, 
when acting on the left and right quantum kinks states, have fractional eigenvalues
 \bea
 S_{\cal L}^z |\pm \frac{1}{4} \rangle_{\cal L} &=& \pm \frac{1}{4}  |\pm \frac{1}{4} \rangle_{\cal L}, \nonumber \\
 S_{\cal R}^z |\pm \frac{1}{4} \rangle_{\cal R} &=& \pm \frac{1}{4}  |\pm \frac{1}{4} \rangle_{\cal R}.
 \label{edgespins}
 \eea
 With these assumptions,   the edge state Hilbert space is given by the tensor product of left and right kinks states (\ref{edgekinks})  and consists into the four states
 \be
 |\pm \frac{1}{4} \rangle_{\cal L} \otimes |\pm \frac{1}{4} \rangle_{\cal R},
 \ee
 which can be sorted out according to the total spin 
 \be
 S^z = S_{\cal L}^z + S_{\cal R}^z,
 \ee
 into two $S^z=0$ states 
 \bea
  | + \frac{1}{4} \rangle_{\cal L} \otimes |- \frac{1}{4} \rangle_{\cal R}, \; |- \frac{1}{4} \rangle_{\cal L} \otimes |+ \frac{1}{4} \rangle_{\cal R}, 
 \label{spinzero} 
 \eea
 and two  $S^z=\pm 1/2$ states
 \bea
  | +\frac{1}{4} \rangle_{\cal L} \otimes |+ \frac{1}{4} \rangle_{\cal R}, \; |- \frac{1}{4} \rangle_{\cal L} \otimes |- \frac{1}{4} \rangle_{\cal R}. 
   \label{spinpm1/2}
 \eea
 Although the left and right edge states carry fractionalized spins $\pm 1/4$ the total spins of the four ground states have integer and half integer values as it should.

 \subsubsection{Majorana Modes and Symmetry Fractionalization}
 
The above analysis matches the  mean field 
approaches \cite{Wen, Keselman2018} based on coupled Majorana chains,
 that we  shall label "x" and "y" chains. It leads 
 the  existence of four local Majorana  modes (two  at each edge) 
\bea
(\sigma^x_{\cal L}, \sigma^y_{\cal L}) \; {\rm and}\; 
(\sigma^x_{\cal R}, \sigma^y_{\cal R}),
\label{Majorana}
 \eea
 which satisfy the Clifford algebra, i.e:  
$\{\sigma^{\mu}_{r}, \sigma^{\nu}_{r'}\}=2  \delta_{r r'}\delta^{\mu \nu}$ where $(\mu, \nu)=(x,y)$ and $(r, r')=(\cal L, \cal R)$. At each edge, the low energy Hilbert space is described by the two $SO(2)_{(\cal L, \cal R)}$  {\it spinors} (\ref{edgekinks}) which span two  irreducible projective representations of  the $U(1)$ symmetry group generated by
\be
s_{(\cal L, \cal R)}=\frac{1}{4 i} [ \sigma^x_{(\cal L, \cal R)}, \sigma^y_{(\cal L, \cal R)}  ].
\ee
These two representations are eigenvectors of the local fermionic parity operators
 \bea
 {\cal P}_{\cal L} &=& -i \sigma^x_{\cal L}  \sigma^y_{\cal L} \ = \sigma^z_{\cal L}, \nonumber \\ 
 {\cal P}_{\cal R} &=&  -i \sigma^x_{\cal R}  \sigma^y_{\cal R} = \sigma^z_{\cal R},
 \label{fparitiesLR}
 \eea
with $[{\cal P}_{\cal L}, {\cal P}_{\cal R}]=0$ and ${\cal P}_{(\cal L, \cal R)}=\pm 1$. 
They  are related to the local spin operators by 
\be
\sigma^z_{(\cal L, \cal R)} =  -i e^{i 2 \pi S^z_
 {(\cal L, \cal R)}},
\label{spisigma}
\ee
or equivalently by $S^z_{(\cal L, \cal R)} =   \frac{1}{4} \sigma^z_{(\cal L, \cal R)}$, so that   the states 
$|\pm \frac{ 1}{4}\rangle_{(\cal L, \cal R)}$ have
local fermionic parities ${\cal P}_{(\cal L, \cal R)}=\pm 1$.
One may verify, using  (\ref{Nevenodd}), that the total fermionic parity operator fractionalizes into
\be
{\cal P} = -{\cal P}_{\cal L} {\cal P}_{\cal R} = (-1)^N.
\label{fparity}
\ee
In this scheme, the two $S^z=0$ states in (\ref{spinzero}), with an even total fermionic parity ${\cal P}=+1$ are the tensor products of states with opposite local parities. The two states (\ref{spinpm1/2}) with total spins $S^z=\pm 1/2$ have an odd total fermionic parity ${\cal P}=-1$ and are the tensor products of states with the same local parities.

When projected onto the low energy subspace of the edges kinks, the $\mathbb{Z}_2$  symmetry 
group (\ref{z2}) also  fractionalizes between the two edges into local   $\mathbb{Z}_{2,(\cal L, \cal R)}$ symmetry groups 
\be
\mathbb{Z}_{2,(\cal L, \cal R)}=\{1, \sigma^x_{(\cal L,\cal R) }\}, \; (\sigma^x_{(\cal L,\cal R)})^2 = 1. 
\label{z2LR}
\ee
The local spin flip operators $ \sigma^x_{(\cal L,\cal R) }$  reverse the spins $S^z_{(\cal L, \cal R)}$ (as well as  the fermion  parities ${\cal P}_{(\cal L, \cal R)}$) at each edge and   the total spin flip
operator $\tau$ of the $\mathbb{Z}_2$  symmetry   in (\ref{z2}) is given by
 \bea
 \tau =  i \sigma^x_{\cal L}\sigma^x_{\cal R}.
 \label{taudiag}
 \eea
 Since $[\tau, {\cal P}]=0$,  in each fermionic parity sector ${\cal P}=\pm$, one may sort out the  states (\ref{spinzero}) and (\ref{spinpm1/2}) into symmetric  and antisymmetric states  $|{\cal P}=\pm,  \tau = \pm \rangle$ with respect
 to the total spin flip operator $\tau$, i.e:
 \be
 \cal P |\pm, \tau \rangle = \pm |\pm, \tau \rangle, \;  
 \tau |\cal P, \pm \rangle = \pm |\cal P, \pm \rangle,
 \label{symasym}
 \ee
 where 
 \bea
 |+, \pm \rangle &=& \sqrt{\frac{1}{2}}\left(  | + \frac{1}{4} \rangle_{\cal L} \otimes |- \frac{1}{4} \rangle_{\cal R} \pm |- \frac{1}{4} \rangle_{\cal L} \otimes |+ \frac{1}{4} \rangle_{\cal R}\right), \nonumber \\
 |-, \pm \rangle &=&  \sqrt{\frac{1}{2}} \left(| +\frac{1}{4} \rangle_{\cal L} \otimes |+ \frac{1}{4} \rangle_{\cal R} \pm |- \frac{1}{4} \rangle_{\cal L} \otimes |- \frac{1}{4} \rangle_{\cal R}\right).\nonumber \\
 \label{Ptaupm}
 \eea
 
 While working in this basis it suitable to introduce new Majorana operators, $\lambda_{(\cal L,\cal R) }$ and  $\eta_{(\cal L,\cal R) }$, which are associated with the two commuting
$\mathbb{Z}_2$ symmetries of the problem: the $\mathbb{Z}_2 =\{1, \tau\}$  symmetry   (\ref{z2})  associated with the spin flip symmetry of the Hamiltonian and the $\mathbb{Z}^F_2 =\{1, \cal P\}$  fermion parity symmetry. With the correspondance
\bea
\lambda_{\cal L } &=& \sigma^z_{\cal L },\ \ \ \  \lambda_{\cal R }  \   = -i  \sigma^y_{\cal L }\sigma^x_{\cal R }, \nonumber \\
\eta_{\cal L } &=& \sigma^z_{\cal L } \sigma^x_{\cal R }, \ \eta_{\cal R} = \sigma^y_{\cal R },
\label{lambdaeta}
\eea
we check that $[\lambda_r, \eta_{r'}]=0$, $\{\lambda_r, \lambda_{r'}\} = \{\eta_r, \eta_{r'}\}= 2 \delta_{r r'}$, and
\be
\tau= i \lambda_{\cal L }\lambda_{\cal R },\ {\cal P}= i \eta_{\cal L }\eta_{\cal R }.
\label{z2z2F}
\ee
The problem then decouples into two commuting $\lambda$ and $\eta$ Majorana modes which are associated with the $\mathbb{Z}_2$ and $\mathbb{Z}^F_2$ symmetries respectively. In a given total fermion parity sector the Majorana $\lambda_{(\cal L,\cal R) }$   exchange symmetric and antisymmetric states $|\cal P, \cal \tau \rangle \leftrightarrow
|\cal P, -\cal \tau \rangle$ while the Majorana $\eta_{(\cal L,\cal R) }$ reverse the fermion parity of either 
the symmetric or the antisymmetric states $|\cal P, \cal \tau \rangle \leftrightarrow
|-\cal P, \cal \tau \rangle$.

\subsubsection{Particle Number Conservation}

So far we have described the spin sector only. Including the charge degrees of freedom is the equivalent 
to enforce particle number conservation together with the constraint (\ref{Nevenodd}).
In a system with overall conservation of the total number of fermions $N$, the states with 
different fermionic parities $\cal P$ must differ by an odd number of fermions. Hence the minimum 
energy difference between the states with ${\cal P}=+1$ and  ${\cal P}=-1$ in (\ref{Ptaupm}) is given
by the energy cost of adding or removing a charge in the system. This is the charging energy which,
in a Luttinger liquid, goes to zero as $1/L$ in the thermodynamical limit. Thus the four states  (\ref{spinzero}, \ref{spinpm1/2}) or equivalently  (\ref{Ptaupm}) are degenerate in the $L \rightarrow \infty$ limit. 
At finite size though, the effect  of particle number conservation  is to lift the four-fold degeneracy of the edge states.  In a finite system the ground state of the system is (to the exponential accuracy in the system size) only doubly degenerate. The states with opposite fermion parities $|\pm \cal P, \cal \tau \rangle$ are separated by a  gap of order $1/L$ whereas,  in each parity sector,  the symmetric and antisymmetric states $| \cal P, \pm \cal \tau \rangle$ are (quasi-)degenerate with a much smaller  energy splitting $\delta E \sim e^{-mL}$. The  resulting twofold (quasi-)degeneracy is exhausted by the two zero energy Majorana modes  $(\lambda_{\cal L},\lambda_{\cal R})$ (see Eqs. (\ref{lambdaeta}) ) which are localized at each edge of the system and confer the STS phase a topological degeneracy which results from the
$\mathbb{Z}_2$ symmetry of the problem.

 Strictly speaking  the arguments leading to the existence of  the symmetry protected  zero energy edge modes (\ref{edgekinks}), and concurrently   to the topological degeneracy, are valid in the semi-classical regime which corresponds to the strongly anisotropic regime of the $U(1)$ Thirring model. Hence, the mere existence of a stable topological phase that extends in the whole STS phase rely on the assumption that  quantum fluctuation do not spoil the nice picture described above. This is to be particularly true in the weak coupling regime, i.e: $0 < g_{\parallel} \ll 1 , 0 < -g_{\perp} \ll 1$, where  quantum fluctuations are strong. We also stress that at the heart of  the existence of {\it fractional} edge states (\ref{edgekinks}) is the assumption that the fractionalization of the spin quantum number at the edges is a genuine phenomenon. This implies that the fractional spin operators $S_{(\cal L, \cal R)}^z$, defined semi-classically in (\ref{SzL},\ref{SzR}),  have to be given a proper  sense in the full quantum theory as  sharp quantum observable with zero variance in the thermodynamical limit, i.e:
 \be 
\lim_{ L \rightarrow \infty} \langle (S_{(\cal L(\cal R)}^z)^2 \rangle -  (\langle S_{(\cal L,\cal R)}^z \rangle)^2 = 0.
\label{zerovariance}
\ee 
In a strongly interacting theory this is a highly non trivial issue. It is only, to our knowledge,  for free massive fermions interacting with a classical   soliton-anti-soliton field, that  the  fractionalization at the fermion charge at the locations of the soliton and anti-soliton fields has been clearly established \cite{Jackiw}.  In the present context, which  corresponds to  $\beta^2 = 4 \pi$,
 the soliton  and the anti-soliton can be seen as the two left and right boundaries and  the fractionalized fermion charge $\pm 1/2$ 
 corresponds to the fractional  spin $\pm 1/4$ quantum edge states. 
This  supports the results obtained above  in semi-classical analysis. 
However, it says nothing about quantum regime ($\beta^2 > 4 \pi$)  where quantum fluctuations are strong. To investigate this regime other methods are required. This will be provided in the next
section when solving the model (\ref{Hamiltonian}) using the Bethe Ansatz.

 \section{Open Boundaries: The Bethe Ansatz Solution}
\label{sec:betheAnsatz}

In this section we solve exactly, using the Bethe Ansatz, the $U(1)$ Thirring model when OBC are imposed on the fermions.  We shall present the Bethe equations for all values
of the couplings $g_{\parallel}$ and $g_{\perp}$. In the following we shall present our results for the regions A($\widehat{A}$)
of the phase diagram (see Fig. (\ref{fig:PD}))
and on the GN($\widehat{GN}$) lines at  $g_\parallel= \pm g_\perp$   where universal answers can be obtained in the scaling   limit. In the latter  limit, the cutoff $D$ is taken to  infinity while the mass (the spin gap)  $m$ is kept  fixed. This  corresponds  to  the region of small  couplings $(|g_{\parallel|}, |g_{\perp}| )\ll 1$ where $\beta^2\sim 8\pi$  in the SG model (\ref{sineGordon}) and this is precisely  the quantum regime we want to investigate. In the regions C($\widehat{C}$),
as well known, taking the scaling limit is a non trivial issue since, as seen in Fig. (\ref{fig:PD}),  the limit of infinite cutoff corresponds to a strong coupling fixed point (the theory is not asymptotically free) which nature is, to our knowledge, unknown. In the following we shall thus concentrate on the portions of both the SSS and STS phases that correspond to the regions A($\widehat{A}$)  and the GN($\widehat{GN}$) lines in the Fig. (\ref{fig:PD}).

\subsection{Overview of Bethe Ansatz solution}

Our main focus will be on the the effects of the open boundaries on the ground state properties and we shall not discuss their effects on the  gapped excitations. Before proceeding to the actual calculation we present here an outline of the results. We shall show that the model when defined on a line segment with open boundary conditions is integrable and that its properties are given by a set of algebraic equations, the Bethe Ansatz equations, which in addition to incorporating the scattering dynamics of the model also  incorporate the boundary conditions. We shall derive the equations and discuss in detail the boundary effects  that follow from the presence of the boundary terms in the Bethe Ansatz equations.  The solutions of the equations, the Bethe roots, together with the total spin $S^z$ usually characterize the eigenstates of the Hamiltonian, its ground state in particular (see below).

\smallskip

Analyzing the equations we shall find that in the SSS phase the ground state is unique with a total spin $S^z=0$ as in the periodic boundary conditions case \cite{ Japaridze}. In this state all the Bethe roots are real. In the STS phase on the other hand  we find \textit{three} ground states which are degenerate in the thermodynamic limit. Two of the states, denoted $|\widehat{\frac{1}{2}}\rangle$ and $|-\widehat{\frac{1}{2}}\rangle$,  have spins $S^z=1/2$ and $S^z=-1/2$ respectively. 
The state with spin $S^z=1/2$ is constructed from  Bethe reference state with all spin up and the state  with spin $S^z=-1/2$ is constructed from Bethe reference state with all spin down. These states have all real Bethe roots and have identical Bethe root distribution.
The third state $|\widehat{0}\rangle$ has spin $S^z=0$. It is constructed  by adding a purely imaginary solution to either of the states $|\widehat{\frac{1}{2}}\rangle$, $|-\widehat{\frac{1}{2}}\rangle$. Purely imaginary Bethe roots are referred to as boundary strings and correspond to boundary bound states \cite{skorik}. 

So far our Bethe Ansatz solution for symmetric OBC  appears to disagree with the semi-classical predictions which predicts two $S^z=0$ states. We note however that in the presence of symmetries the solutions to Bethe equations might not give all the states in the Hilbert space. To obtain these states, one needs to apply operators associated with the symmetries to the states obtained directly from the Bethe equations - examples are spin raising operators applied to a highest weight state obtainable as Bethe Ansatz state so as to complete  a $SU(2)$ multiplet. 

In the present case the space parity symmetry $x\rightarrow -x$ induces two equal boundary terms in the Bethe equations of the STS phase, due to which the boundary string  occurs as a double pole. One expects that this double pole corresponds to two states, namely, there exists another singlet state $|\widehat{0}'\rangle$ in addition to the state $|\widehat{0}\rangle$. As this state $|\widehat{0}'\rangle$ cannot be obtained directly from the Bethe equations and the construction of a corresponding generating symmetry operator in the Bethe Ansatz framework is a non trivial task, we resolve this issue  by considering a slightly \textit{asymmetric} boundary conditions  allowing a small twist $\epsilon'$  between the left and right moving fermions at the right boundary (to be defined in \eqref{twist1}) . This splits the double pole giving rise to the expected additional boundary string solution, the fourth state $|\widehat{0}'\rangle$, which is  obtained in the limit where $\epsilon'\rightarrow 0$.

This asymmetric BC which also break the $\mathbb{Z}_2$ symmetry do not change the number of ground states in the SSS phase which remains  unique with spin $S^z=0$ in the scaling limit. In the STS phase we obtain four states, two with spin $S^z=\pm 1/2$ and two with spin $S^z=0$ in the scaling limit. As before,  the state $|\widehat{\frac{1}{2}}\rangle_{\epsilon'}$  with spin $S^z=1/2$ is constructed from Bethe reference state with all spin up and the state $|-\widehat{\frac{1}{2}}\rangle_{\epsilon'}$,  with spin $S^z=-1/2$, is constructed from Bethe reference state with all spin down. These two states have all real roots but have now slightly different Bethe root distributions and they differ in their energy: $E_{S^z=1/2}-E_{S^z=-1/2}=\delta E_{\epsilon^\prime}$. Again, a state $|\widehat{0}\rangle_{\epsilon'}$ with spin $S^z=0$ can be obtained by adding a purely imaginary solution  to the state with spin $S^z=1/2$ and is degenerate with it. However,  due to the presence of the asymmetry another purely imaginary solution exists,  which when added to the state with spin $S^z=1/2$ gives a 
 state $|\widehat{0}'\rangle_{\epsilon^\prime}$  with spin $S^z=0$
which is degenerate with the state with spin $S^z=-1/2$.

\begin{table}[h]
\centering
\caption{Total spin and energy in the scaling limit of low lying states in the STS phase with asymmetric boundary conditions}
\begin{tabular}{ccc}
\hline
\hline
  State  & \;\;Total spin &\;\; Energy \\
\hline
 $|\widehat{0}'\rangle_{\epsilon^\prime}$ & 0       & 0      \\
 $|\widehat{-\frac{1}{2} }\rangle{\epsilon^\prime}$& -1/2  & 0    \\
  $|\widehat{0}\rangle_{\epsilon'} $ & 0  & $\delta E_{\epsilon^\prime}  $ \\
$|\widehat{\frac{1}{2}}\rangle{\epsilon^\prime}$ & 1/2 &$\delta E_{\epsilon^\prime}$\\
\hline
\hline
\end{tabular}
\end{table}

In the limit $\epsilon^\prime\rightarrow 0$, when the asymmetry vanishes,   the energy difference $\delta E_{\epsilon^\prime}\rightarrow 0$ and the states $|\widehat{\frac{1}{2}}\rangle_{\epsilon'}$, $|-\widehat{\frac{1}{2}}\rangle_{\epsilon'}$ transform into states $|\widehat{\frac{1}{2}}\rangle$ and $|-\widehat{\frac{1}{2}}\rangle$ respectively. The two singlet states $|\widehat{0}\rangle_{\epsilon^\prime}$ and $|\widehat{0}'\rangle_{\epsilon^\prime}$ yield two different spin singlet states $|\widehat{0}\rangle$ and $|\widehat{0}'\rangle$ which are quasi degenerate in the limit $\epsilon'\ll 1$. The state $|\widehat{0}'\rangle$ is precisely the state we wished to construct in the symmetric BC case. We thus conclude that the fourfold ground state degeneracy found in the semi-classical analysis by Kesselman and Berg \cite{Keselman2015} in the strongly anisotropic regime ($g_{\parallel} \gg 1$) survives strong quantum fluctuations down to  weak couplings  in the region $\widehat{A}$ and on the dual Gross-Neveu line $\widehat{GN}$.

\subsection{Bethe Equations}

Since the  Hamiltonian (\ref{Hamiltonian}) commutes with total particle number $N$ (\ref{number}), $\mathcal{H}$ can be diagonalized by constructing the exact eigenstates in each $N$ sector. From here on, for notational convenience,  we shall use the notation  $(+, -)$  to indicate the chirality index of the fermions replacing $(R, L)$ notation. The $N$-particle eigenstate takes the standard reflection Bethe Ansatz form of a plane wave expansion in  $N!\, 2^N$ different regions of coordinate space. The state is labeled by momenta $k_j, j=1\cdots N$, the same in all regions,  and  is given by,
\bea\label{nparticle}
\ket{\{k_j\}}=
\sum_{Q,\vec{a},\vec{\sigma}}\int \theta(x_Q) A^{\{\sigma\}}_{\{a\}}[Q] \prod_j^N e^{i\sigma_j k_jx_j}\psi^{\dagger}_{a_j\sigma_j}(x_j)\ket{0}.
\eea
with energy eigenvalue $E=\sum_j k_j$.  In the above equation, the  sum is to be taken  over all  spin and chirality configurations specified by $\{a\}=\{a_1\dots a_N\}$, $\{\sigma\}=\{\sigma_1\dots \sigma_N\}$ as well as different orderings of the $N$ particles. These different orderings correspond to elements of the symmetric group $Q\in \mathcal{S}_N$. Here  $\theta(x_Q)$ denotes the Heaviside function which is nonzero only for that particular ordering $Q$. The amplitudes $A^{\vec{\sigma}}_{\vec{a}}[Q]$ are related to each other by the various $S$-matrices. Amplitudes which differ by changing the chirality of the rightmost and leftmost particle are related by the boundary S-matrices which  are identities in our system owing to the open boundary conditions \eqref{OBC}. Amplitudes which are related by swapping the order of particles with different chiralities are related by the particle-particle $S$-matrix, which is given by, \cite{duty}
\bea S^{ij}= \left(\begin{array}{cccc} 1&&&\\&\displaystyle{\frac{\sinh(f)}{\sinh(f+\eta)}}&\displaystyle{\frac{\sinh(\eta)}{\sinh(f+\eta)}}&\\&\displaystyle{\frac{\sinh(\eta)}{\sinh(f+\eta)}}&\displaystyle{\frac{\sinh(f)}{\sinh(f+\eta)}}&\\&&&1\end{array}\right),\eea
where $\eta=-iu$ and $f$, $u$ are related to $g_{\parallel}$ and $g_{\perp}$ through the relations \bea\cos(u)=\frac{\cos(g_{\parallel})}{\cos(g_{\perp})}, \hspace{2mm} \frac{\sin(u)}{\tanh(f)}=\frac{\sin(g_{\parallel})}{\cos(g_{\perp})}.\eea An additional  $S$-matrix, denoted by $W^{ij}$, is also required. It relates amplitudes that differ by exchanging particles of the same chirality. This is given by 
\bea
W^{ij}=P^{ij}.
\eea
 The consistency of the solution is then guaranteed as the $S$- and $W$ matrices satisfy the Yang-Baxter and Reflection equations \cite{Sklyannin, Cherednik, ZinnJustin}. 
 
 Imposing the boundary condition at $x=\pm L/2$  quantizes the single particle momenta $k_j$ which are expressed in terms of $M$ parameters $\lambda_\beta$, the Bethe rapidities or Bethe roots, which  satisfy a set of coupled nonlinear equations called the Bethe equations. In a state, $M$ denotes the number of down spins and $N-M$ is the number of up spins and vice-versa. We use the method of Boundary Algebraic Bethe Ansatz to obtain the logarithmic form of Bethe equations, which take different forms in different regions of Fig\eqref{fig:PD}.

For definiteness we give the explicit form of 
the Bethe equations in the regions $A$ and $\widehat{A}$ below.

\bea\nonumber
 \sum_{\sigma=\pm}N\Theta(\lambda_\alpha+\sigma f/2u ,1/2)- 2\Theta(\lambda_\alpha+i\tau \pi/2u,1/2)\\ \label{logBae}=\sum_{\beta=1}^{M} \sum_{\sigma=\pm}\Theta\left(\lambda_\alpha+\sigma \lambda_\beta,1\right)+2i\pi I_\alpha,
 \label{BEA}
 \eea

\bea\label{logEnergy}
 k_j=\frac{\pi n_j}{L}+\frac{i}{2L}\sum_{\beta=1}^M\sum_{\sigma=\pm}\Theta(f/2u+\sigma\lambda_\beta,1/2),
\eea
where $\displaystyle{\Theta(x,y)=\log\left(\frac{\sinh(u(x+iy))}{\sinh(u(x-iy))}\right)}$. The second term in Eq.(\ref{BEA}) is a boundary term where $\tau=1$ in the region $A$, and $\tau=0$ in region $\widehat{A}$.  The parameters $f,u$ are real in the regions $A$ and $\widehat{A}$. The $GN$ and $\widehat{GN}$ lines correspond to the isotropic limit $(f, u \rightarrow 0,f/u=1/g)$ of the Bethe equations of regions $A$ and $\widehat{A}$ respectively \cite{Japaridze}. We work in the region where $u<\pi/2$ which corresponds to $4\pi<\beta^2<8\pi$ in SG.

The boundary term in the topological phase leads to a dramatic change in the degeneracy of the ground state in the region $\widehat{A}$ and on the $\widehat{GN}$ line. 

The Bethe roots govern the spin degrees of freedom of the system and $M\leq N/2$ gives the total $z$-component of spin, $S^z=N/2-M$. The solutions to equations of type \eqref{logBae} are well studied in the literature \cite{Takahashi},\cite{ODBA}. The solutions $\lambda_\alpha$ can be real or take complex values in the form of strings. In order to have a non vanishing wavefunction they must all be distinct, $\lambda_\alpha \neq \lambda_\beta$. In addition, the values $\lambda_\alpha=(0,i\pi/2u)$ should also be discarded as they result in a vanishing wavefunction \cite{ODBA}. Bethe equations of the type \eqref{logBae} are reflective symmetric, that is they are invariant under $\lambda_\alpha\rightarrow -\lambda_\alpha$ transformation. Due to this symmetry, solutions to the Bethe equations occur in pairs $\{-\lambda_\alpha,\lambda_\alpha\}$. The integers 
$n_j$ and $I_\alpha$ arise from the logarithmic branch and serve as the quantum numbers of the states. The quantum numbers $I_\alpha$ correspond to the spin degrees of freedom while the quantum numbers $n_j$ are associated with the charge degrees of freedom and they must all be different. $I_\alpha$ and $n_j$ can be chosen independently implying the charge spin decoupling. Minimizing the ground state energy results in a cutoff such that the $\pi|n_j|/L < \pi D$ where $D=N/L$ is the density \cite{Andrei}.  

\subsection{The SSS phase}
\label{sec: SSSGS}

This corresponds to the regions $A$ and the $GN$ line, as displayed in the Fig. \ref{fig:PD}.
We shall consider them separately in the following.

\subsubsection{Region $A$} 

The ground state is given by the particular choice of charge and spin quantum numbers $n^0_j$ , $I^0_\alpha$, where $n^0_j$ are consecutively filled from the lower cutoff $-LD$ upwards, and the integers $I^0_\alpha$  take consecutive values which corresponds to real valued  $\lambda_\alpha$ roots in the region $A$. In the limit $N\rightarrow \infty$ the Bethe roots fill the real line and the ground state can be described by $\rho(\lambda)$ the density of solutions $\lambda$, from which the properties of the ground state can be obtained. Reflection symmetry of the Bethe equations \eqref{logBae} allows us to define $\lambda_{-\alpha}=-\lambda_\alpha, \; \lambda_0=0$ \cite{XXXkondo} and introduce the counting function $\nu(\lambda)$ such that $\nu(\lambda_\alpha)=I_\alpha$. Differentiating \eqref{logBae}, and noticing that $\rho(\lambda)=\frac{d}{d\lambda}\nu(\lambda)$ \cite{trieste}, we obtain the following integral equation, 
\bea
h_{A}(\lambda)&=&\rho_{A}(\lambda)+\sum_{\sigma=\pm}\int_{-\infty}^{+\infty} d\mu \; a_2(\lambda-\sigma\mu)\rho_{A}(\mu),\nonumber \\
\label{deneqn}
\eea
where $\rho_{A}$ stands for the ground state density distribution in the region $A$ and $h_{A}(\lambda)=Na_1(\lambda+\sigma f/2u)+a_2(\lambda)+ a_1(\lambda)-b_1(\lambda)$ where
\be
a_n(x)= \frac{u}{\pi} \frac{\sin(n u)}{\cosh(2ux)-\cos(n u)},
\ee
\be
b_n(x)= -\frac{u}{\pi} \frac{\sin(n u)}{\cosh(2ux)+\cos(n u)}.
\ee
Note that we have excluded the root $\lambda=0$ and also applied the restriction $\lambda_\alpha\neq\lambda_\beta$.

Solving \eqref{deneqn} by Fourier transformation \cite{Doikoucritical} we obtain the Fourier transformed ground state distribution of Bethe roots in the region $A$
\bea
\label{gsdensityA}\tilde\rho_{A}(\omega)=\displaystyle{ \frac{N\cos[\frac{f\omega}{2u}]+\frac{1}{2}\left(\frac{\sinh((\pi-2u)(\omega/2u))+\sinh(\omega/2)}{\sinh((\pi-u)(\omega/2u))}+1\right)}{\cosh[\frac{ \omega}{2}]}}. 
\eea 
The term which is proportional to $N$ corresponds to the bulk
contribution  while  the terms of order $N^{(0)}$ can be associated with the boundaries at $x=(-L/2,L/2)$. The number of Bethe roots $M_{A}$ in the ground state of region $A$  is given by
\bea
2M_{A}+1=\int_{-\infty}^{+\infty}d\lambda \; \rho_{A}(\lambda),
\label{nrootsa}
\eea
 from which the $z$-component of spin $(S^z)_{A}$ of the ground state in this region is  obtained using  the relation $S^z_{A}=N/2-M_{A}$.  Taking into account that  $\tilde\rho(0)=\int\mathrm{d}\lambda\, \rho(\lambda)$ along with \eqref{gsdensityA} we find that in the scaling limit, i.e. when $|g_\parallel |\ll1,|g_\perp|\ll1, u\ll1$, 
 \bea 
 (S^z)_{A}=0.
 \eea 
We thus find from (\ref{Nevenodd}) that the ground state in the region $A$ has an even number of fermions
and hence  an {\it even} fermion parity ${\cal P}=+1$. It is non degenerate and is a $\mathbb{Z}_2$ singlet.

\subsubsection{ $GN$ Line } 

 On the $GN$ line, the Bethe equations can be obtained by taking the limit $f,u\rightarrow 0, f/u=1/g$, which leads to the limit \bea\nonumber\text{log}\left(\frac{\sinh(u(x+iy))}{\sinh(u(x-iy))}\right)&\rightarrow &\text{log}\left(\frac{x+iy}{x-iy}\right),\\\label{isolimit} \text{log}\left(\frac{\cosh(u(x+iy))}{\cosh(u(x-iy))}\right)&\rightarrow1&\eea in \eqref{logBae}. 
 We obtain the following integral equation
 \bea \nonumber h_{GN}(\lambda)=\rho_{GN}(\lambda)+\sum_{\sigma=\pm}\int_{-\infty}^{\infty}d\mu\;\rho_{GN}(\mu)\;\varphi(\lambda-\sigma\mu,1), \\\label{eqgn}\eea
 
 where $\rho_{GN}$ stands for the ground state density distribution on the $GN$ line and $h_{GN}(\lambda)=\sum_{\sigma=\pm}2N\varphi(2\lambda+\sigma/g,1)+2\varphi(2\lambda,1)+\varphi(\lambda,1)$ where \be\varphi(x,a)=(1/\pi)(a^2+x^2)^{-1}.\ee

 Solving \eqref{eqgn} by Fourier transformation \cite{PRA1} we obtain the ground state distribution on the $GN$ line,
\bea
\label{gsdensitygn}\tilde\rho_{GN}(\omega)=\frac{N\cos[\frac{\omega}{2g}]+ \frac{1}{2}+\frac{1}{2}e^{-\frac{|\omega|}{2}}}{\cosh[\frac{\omega}{2}]}.
\eea
The number of Bethe roots is given by an equation similar to \eqref{nrootsa}, using which we obtain 
\bea 
(S^z)_{GN}=0,
\eea
for the ground state in the region $GN$.  Exactly as in the region $A$ in the SSS phase, the ground state on the $GN$ line is non degenerate and has an even fermion parity ${\cal P}=+1$. It is actually, on top of being 
a $\mathbb{Z}_2$ singlet, an $SU(2)$ singlet.

\smallskip

In summary, we have  seen that, although the descriptions of the ground state in terms of the Bethe root distribution is different in the region $A$ and on the $GN$ line, the ground state belongs to the even fermion parity sector
${\cal P}=+1$. It is  non degenerate,  has a zero total spin $S^z=0$ and is at least a $\mathbb{Z}_2$ singlet.
Labeling the SSS ground state as $|0\rangle$ we have
\be
S^z|0\rangle= 0,\ {\cal P}|0\rangle= |0\rangle, \ \tau |0\rangle=|0\rangle,
\ee
where ${\cal P}=(-1)^N$ is the fermionic parity operator and $\tau$ is the total spin flip operator
 generating the $\mathbb{Z}_2$ symmetry group (\ref{z2}).
 As in the case where  periodic boundary condition are imposed, the ground state properties are the same in these
regions and there is no phase transition between them.

\subsection{The STS Phase}
\label{sec: STSGS}

This correspond to the regions $\widehat A$ and the $\widehat{GN}$ line. As already emphasized, in the  STS phase the boundary term in the Bethe equations \eqref{logBae} leads to a change in the ground state degeneracy as we shall now see. 

\subsubsection{Region $\widehat{A}$}

By following the same procedure as before, we obtain the following integral equation
\bea\label{deneqn2}
h_{\widehat{A}}(\lambda)=\rho_{\widehat{A}}(\lambda)+\sum_{\sigma=\pm}\int_{-\infty}^{\infty}d\mu \; a_2(\lambda-\sigma\mu)\rho_{\widehat{A}}(\mu),
\eea
where $\rho_{\widehat{A}}$ stands for the ground state density distribution in the region $\widehat{A}$ and $h_{\widehat{A}}(\lambda)=Na_1(\lambda+\sigma f/2u)+a_2(\lambda)- a_1(\lambda)+b_1(\lambda)$.
Solving \eqref{deneqn2} by Fourier transform we obtain the ground state distribution in region $\widehat{A}$
\bea
\label{gsdensityAt}\tilde\rho_{\widehat{A}}(\omega)=  \frac{N\cos[\frac{f\omega}{2u}]+\frac{1}{2}\left(\frac{\sinh((\pi-2u)(\omega/2u))-\sinh(\omega/2)}{\sinh((\pi-u)(\omega/2u))}-1\right)}{\cosh[\frac{\omega}{2}]}.
\eea
Notice that the second and third term of the boundary contribution in the above expression have opposite sign compared to those in region $A$ (\ref{gsdensityA}). As a consequence,  using  \eqref{nrootsa}, we find that the ground state in region $\widehat{A}$ has a non zero spin
\bea (S^z)_{\widehat{A}}=\frac{1}{2},
\eea
which, from (\ref{Nevenodd}), corresponds to an odd number of particles $N$ and hence has an {\it odd}
fermion parity ${\cal P}=-1$. Due to the $\mathbb{Z}_2$ symmetry (\ref{z2}) we immediately deduce that 
there is another ground state in the same fermion parity sector, degenerate with the above,  which has the opposite spin \bea
(S^z)_{\widehat{A}}=-\frac{1}{2}.
\eea
Actually, this state  can be obtained by choosing the Bethe reference state with all spins  down instead of up \cite{korepin1993quantum}. The two states $S^z=\pm1/2$ have the same Bethe root distribution and transform into each other under the action of the $\mathbb{Z}_2$ generator $\tau$.
This is to be contrasted with the  situation in the region $A$ where the ground state, having $S^z=0$, has an even
fermion parity ${\cal P}=+1$ and  is a $\mathbb{Z}_2$ singlet.

\subsubsection{$\widehat{GN}$ Line}

 The Bethe equations on $\widehat{GN}$ line are rational just as in the case of $GN$. They can be obtained by taking the isotropic limit \eqref{isolimit} of the Bethe equations in region $\widehat{A}$. We obtain the following integral equation
 
 \bea\nonumber h_{\widehat{GN}}(\lambda)=\rho_{\widehat{GN}}(\lambda)+\sum_{\sigma=\pm}\int_{-\infty}^{\infty}d\mu \; \rho_{\widehat{GN}}(\mu)\; \varphi(\lambda-\sigma\mu,1),\\ \label{eqgnt}\eea
 where $\rho_{\widehat{GN}}$ stands for the ground state density distribution on the $\widehat{GN}$ line and $h_{\widehat{GN}}(\lambda)=\sum_{\sigma=\pm}2N\varphi(2\lambda+\sigma/g,1)-2\varphi(2\lambda,1)+\varphi(\lambda,1)$ 
 
 Solving the \eqref{eqgnt} by Fourier transformation we obtain the following distribution of Bethe roots in the ground state on the $\widehat{GN}$ line
\bea\label{gsdensitygnt}\tilde\rho_{\widehat{GN}}(\omega)=\frac{N\cos[\frac{\omega}{2g}]- \frac{1}{2}+\frac{1}{2}e^{-\frac{|\omega|}{2}}}{\cosh[\frac{\omega}{2}]}.\eea
Using equation similar to \eqref{nrootsa} we obtain 
two degenerate ground states with spins
\bea
(S^z)_{\widehat{GN}}=\pm \frac{1}{2}.
\eea
As in the region $\widehat{A}$ these two states have an odd fermion parity
and transform into each other under $\tau$. Notice that, contrarily to the ground state on the $GN$ line,
they are obviously not  $SU(2)$ singlet states. This is consistent with the fact that on $\widehat{GN}$ line
the model is only $U(1)_s \otimes  \mathbb{Z}_2$ symmetric and that the enlarged $\widehat{SU(2)}$ symmetry (\ref{su2dual}) is non local. 

Just as in the  SSS phase, despite having a different descriptions in terms of 
the Bethe root distributions, there is no phase transition between these regions in the STS phase. Here the ground state 
belongs to the {\it odd} fermion parity sector ${\cal P}=-1$ and is doubly
degenerate, each ground state having  spins $S^z=\pm 1/2$. The degeneracy 
here is to be understood as the consequence of the non vanishing of the spin in the ground state and 
of  the $\mathbb{Z}_2$ symmetry which reverses the total spin.
This is to be contrasted with what happens in the  SSS phase where the ground state, having
a spin zero, is not degenerate.

As we shall see, besides  the two degenerate
states with $S^z=\pm1/2$, there is one more state with  $S^z=0$ which is degenerate 
with the ground states  in the large system size limit $L\rightarrow \infty$. 
This state is a solution of the Bethe equations that involve
a boundary string which corresponds to a boundary mode.

 \subsection{Bulk Excitations}
 
 Excitations correspond to states whose quantum numbers, $n_j$ or $I_\alpha$ have been modified from their ground state configurations. Note that we can choose $n_j$ and $I_\alpha$ independently, meaning that the spin and charge degrees of freedom are decoupled\cite{Haldane, Witten79, Andrei}. In the charge sector the excitations are constructed by removing a number, $n^h<0$ from the sequence $n^0_j$ and adding an extra $n^p>0$. The energy of this excitation is $\delta E=2\pi(n^p-n^h)/L>0$. Gapless excitations such as this are known as holons.
The structure of excitations in the spin sector is more complicated as they arise from  solutions to the Bethe Ansatz equations \eqref{logBae}
for non ground state configurations of the $I_\alpha$ quantum numbers. The lowest energy spin bulk excitation is of two spinons which is constructed by removing two arbitrary Bethe roots, $\lambda^h_1,\lambda^h_2$ from the ground state distribution \cite{Andrei}. Each hole corresponds to a single spinon with spin $+1/2$. The energy of this excitation in all the regions except on the $GN$ and $\widehat{GN}$ lines is \bea\label{tripletenergy}\delta E=  \sum_{l=1}^2D\arctan\left[\frac{\cosh(\pi \lambda^h_l/u)}{\sinh\left(f\pi/2u\right)}\right].\eea
From this we find that the system has dynamically generated a superconducting mass gap in the spin sector  \bea\label{spgb}m=D\arctan{[\sinh(f\pi/2u)]^{-1}}.\eea

\smallskip
{\it Universality:}  Having obtained a dynamically generated mass gap
we may remove the cutoff $D$ and obtain universal answers, in other words  taking the scaling limit, $D\to\infty$ while holding the physical mass $m$ fixed. This corresponds to $g_\parallel \ll 1,g_\perp \ll 1$ or $u\ll 1$. In this limit we have that $m=2De^{-f\pi/2u}$ and the excitation energy of a single spinon becomes $\varepsilon(\lambda)=m\cosh{(\lambda)}$, where $\pi/u$ is absorbed into $\lambda$. On the $GN$ and $\widehat{GN}$ lines, the excitation energy and the mass gap is obtained by the replacement $f/u\rightarrow 1/g$.

\subsection{Boundary Excitations}

The boundary modes arise as purely imaginary solutions of the Bethe equations. These purely imaginary Bethe roots, which correspond to the bound states, appear as  poles in the dressed or physical boundary S-matrix  \cite{gosh,gosz,skorik,mez}.  By observation we  see that, in the limit $N\rightarrow\infty$, the Bethe equations \eqref{logBae} have a {\it unique} solution 
\be
\lambda=\pm i/2,
\label{boundstring}
\ee 
as the two $\pm$ strings  leads to the same state by reflection symmetry. 
This is to be true both in  the region $\widehat{A}$ and on the $\widehat{GN}$ line. Adding the boundary string to either the  $S^z= 1/2$ or the $S^z= -1/2$ ground state  a unique state is found. The reason is that two Bethe states are equivalent if they are described by the same root distribution and in addition have the same total spin $S^z$ values. As already mentioned, both the states with $S^z=\pm1/2$ have the same root distribution. The states obtained by adding the boundary string solution to these states with $S^z=\pm 1/2$ will again have the same root distribution. It turns out that these resulting states both have $S^z=0$ and hence they both are equivalent.

 \subsubsection{Region $\widehat{A}$} 
 
Adding the boundary string \eqref{boundstring} to the Bethe equations in region $\widehat{A}$ \eqref{logBae} results in the following equation \bea\nonumber
-2i\pi I_\alpha+ \sum_{\sigma=\pm}N\Theta(\lambda_\alpha+\sigma f/2u ,1/2)- 2\Theta(\lambda_\alpha,1/2)\\ \label{logBaeb}=\Theta(\lambda_\alpha,1/2)+\Theta(\lambda_\alpha,3/2)+\sum_{\beta=1}^{M-1} \sum_{\sigma=\pm}\Theta\left(\lambda_\alpha+\sigma \lambda_\beta,1\right).
 \label{BEAb}
 \eea

The above equation can be solved by following the same procedure as in the ground state. We obtain the following distribution of Bethe roots 
 \bea
 \tilde{\rho}^b_{\widehat{A}}=\tilde{\rho}_{\widehat{A}}+ \Delta\tilde{\rho}^b_{\widehat{A}}, 
  \label{BetherootsBS}
 \eea
 where $\tilde{\rho}_{\widehat{A}}$ is the ground state  distribution given by (\ref{gsdensityAt}) and the shift
 \be
 \Delta\tilde{\rho}^b_{ \widehat{A}}=-\frac{\sinh((\pi-2u)(\omega/2u))}{\sinh((\pi-u)(\omega/2u))}, 
\ee 
 is due to the presence of the boundary string. In the presence of  the boundary string, the relation between the number of Bethe roots and the density distribution also  takes a different form as compared to \eqref{nrootsa}. Namely
\bea 
\label{nct}2M^b_{\widehat{A}}-1=\int_{-\infty}^{+\infty} d\lambda\; \rho^b_{\widehat{A}}(\lambda),
\eea
from which, using $(S^z)^b_{\widehat{A}}=N/2-M^b_{\widehat{A}}$,  we find, in the scaling limit $u \ll 0$,  the  spin of  this state 
\bea 
(S^z)^b_{\widehat{A}}=0. 
\eea 
 Thus the resulting state corresponding to the boundary string (\ref{boundstring}) is a spin singlet which has fermion parity 
${\cal P}=+1$. From the analysis of the Bethe equations of XXZ spin chain \cite{skorik} with equal boundary terms, we expect  that the wave function associated with this unique fundamental boundary string is exponentially localized near both the left and right  boundaries. Furthermore it is symmetric upon the exchange  of  both boundaries or under space parity $x\rightarrow -x$. However since it has a total spin $S^z=0$, we cannot infer how this state transforms under the $\mathbb{Z}_2$ symmetry  (\ref{z2}). As we shall now see the situation is similar on the $\widehat{GN}$ line.

 \subsubsection{$\widehat{GN}$ line}
 
  On the $\widehat{GN}$ line we find that the addition of the boundary string leads to the following change in the distribution of the Bethe roots 
\bea \tilde{\rho}^b_{\widehat{GN}}=\tilde{\rho}_{\widehat{GN}}+\Delta\tilde{\rho}^b_{ \widehat{GN}}, \hspace{2mm} \Delta\tilde{\rho}^b_{ \widehat{GN}}=-e^{-|\omega|/2}. 
\eea
Using an equation similar to \eqref{nct} we find in the scaling limit
\bea 
(S^z)^b_{\widehat{GN}}=0.
\eea
Again, as expected, we obtain a unique state with $S^z=0$ by adding the boundary string to either of the ground states with spins $\pm 1/2$ on the $\widehat{GN}$ line.

\subsubsection{Boundary String Energy}
\label{sec:bsenergy}

As seen, the addition of the boundary string to the ground 
state with either spins $S^z=\pm 1/2$ in the STS phase leads,  in each of the regions $\widehat{A}$  and $\widehat{GN}$, to a {\it single} new state with spin $S^z=0$ that includes a boundary excitation. To get the energy of this
state, or of the boundary string, we notice that it is given 
by the energy difference, up to chemical potential, between  the ground states with $S^z=0$ and $S^z=\pm 1/2$
\begin{eqnarray}
\label{ebk}E_B=E_{N}-\frac{1}{2}(E_{N-1}+E_{N+1}).
\end{eqnarray}
Here $E_{N}$ refers to the energy of the state with odd number of particles which, in our system, corresponds to the ground states in the topological phase with spin $S^z=\pm 1/2$. Similarly $E_{N+1}$ and  $E_{N-1}$ refer to the energies  of the states with an {\it even} number of particles and spin  $S^z=0$. The latter states   include the added    boundary string. The expression \eqref{ebk} is defined in \cite{Keselman2018} as the binding energy, which precisely measures the energy cost of adding an electron to the system. Where it is shown that this is equal to only the charging energy in the topological phase and is equal to the mass gap in the topologically trivial phase.

As it can be shown, the value of  the boundary string energy  is the same in both the region $\widehat{A}$ and on the $\widehat{GN}$ line.
We shall consequently  evaluate $E_B$ in the region $\widehat{A}$. To this end  
we use \eqref{logEnergy}, from which we obtain the following expression for total energy of a state with $N$  fermions
 \bea
 E=\sum_{j=1}^{N}\frac{\pi}{L}n_j+\frac{iD}{2}\sum_{\sigma=\pm}
 \int_{-\infty}^{\infty}d\lambda\; \Theta\left(\frac{f}{2u}+
 \sigma\lambda,\frac{1}{2}\right)\rho_{\widehat{A}}(\lambda). \nonumber \\
 \label{toten}
 \eea
  From  \eqref{ebk} we find that $E_B$ has two contributions, one from the charge degrees of freedom 
 and one from the spin degrees of freedom: 
 $E_B=E_{\text{charge}}+E_{\text{spin}}$. The charge contribution is given by the charging energy
 \begin{eqnarray}
  E_{\text{charge}}=\sum_{j=1}^{N} \frac{\pi}{L}n_j-\frac{1}{2}\left(\sum_{j=1}^{N+1} \frac{\pi}{L}n_j+\sum_{j=1}^{N-1} \frac{\pi}{L}n_j\right). \nonumber \\
  \label{totenc}
 \end{eqnarray}
 Note that the the charge quantum numbers take all the values from the cutoff $-DL$ upwards. In the ground state with $S^z=\pm 1/2$ they fill all the slots from $n_j=-N \; \text{to}\; n_j=-1$. In the state with one extra particle they fill all the slots from $n_j=-N \;\text{to}\; n_j=0$. In the state with one less particle there is an unfilled slot at $n_j=-1$ which corresponds to a holon excitation. Hence we obtain
 \bea \label{bounden}E_{\text{charge}}=-\frac{\pi}{2L}.\eea

 The spin contribution is given by the expression
 \bea
 E_{\text{spin}}=E_{0}+\frac{iD}{2}\sum_{\sigma=\pm}\int_{-\infty}^{+\infty} d\lambda\; \Theta\left(\frac{f}{2u}+\sigma\lambda,\frac{1}{2}\right)\Delta\rho^b_{A}(\lambda), \nonumber \\
 \label{totens}
 \eea
 where $E_{0}=\frac{iD}{2}\;\Theta(f/2u,1)$ and $\Delta\rho^b_{A}(\lambda)$ is the shift of the Bethe roots distribution due  to the boundary string which is given in (\ref{BetherootsBS}). Evaluating (\ref{totens})  we find that 
 the spin part of the energy of the boundary string is exactly zero in the thermodynamic  limit. Hence this corresponds to a zero energy boundary bound state localized at the two ends of the system in a finite system, the boundary string, which is a solution to the Bethe equations in the limit $N\rightarrow\infty$,  
 have corrections of the order $1/N$. Since finite size  corrections to  the Bethe equations are generally  expected to be  exponentially small in the system size we expect that  $E_{\text{spin}} \sim e^{-m L}$ also. As a result, we thus find that
 the energy of the boundary string is, to the exponential accuracy in the system size, 
 given by the charging energy (\ref{totenc})
 \be
 E_B=-\frac{\pi}{2L} 
 \ee
 and hence vanishes in the thermodynamical limit. 
 
 We thus find  that in the regions $\widehat{A}$ and $\widehat{GN}$ of the STS phase the ground state is only {\it threefold} degenerate in the limit of infinite size
 in contrast with the  {\it fourfold} degeneracy predicted by the semi-classical analysis of the preceding section  (\ref{sec:boundary}). The three ground states  in the STS phase are given by the two $S^z=\pm 1/2$ ground states found in the odd fermion parity sector (\ref{sec: STSGS})  plus a single  $S^z=0$ state in the even fermion parity sector which is obtained from them by adding the boundary string $\lambda=\pm i/2$. Labelling the ground states in the STS phase by their spins 
  \be
 |-\widehat{\frac{1}{2} }\rangle,\; |+\widehat{\frac{1}{2} }\rangle,  |\widehat{0} \rangle,
 \label{STSGSBA}
 \ee
 with
 $
 S^z|\pm \widehat{\frac{1}{2}} \rangle = \pm \frac{1}{2} |\pm \widehat{\frac{1}{2} }\rangle$ and  $ S^z|\widehat{0} \rangle=0
 $,
 we have 
 $
 {\cal P} |\pm \widehat{\frac{1}{2}} \rangle  = -|\pm \widehat{\frac{1}{2} }\rangle$ and $ {\cal P} |\widehat{0}\rangle = |\widehat{0}\rangle.
 $
However, although we clearly have
 $
 \tau |\pm \widehat{\frac{1}{2}} \rangle = |\mp\widehat{\frac{1}{2}} \rangle
  $,
 we cannot infer from our analysis
 whether the state $|\widehat{0}\rangle$ is symmetric  or antisymmetric under   the $\mathbb{Z}_2$ symmetry group generator $\tau$ (\ref{z2}). In either case, when comparing
  the semi-classical prediction given in  Eq.(\ref{symasym}) to Eq.(\ref{STSGSBA}), one notes that one extra spin singlet state  is  expected but not  obtained from the solution to the Bethe equations.  One may then wonder whether the  fourfold degeneracy predicted in the semi-classical approximation survives into the full quantum regime.
We shall argue in the following that this is  the case. 
 
\subsection{Asymmetric Boundary Conditions}

 Actually, in the presence of symmetries
 not all states are given as solutions of the Bethe Ansatz equations. A well known example
 is  the bulk spin one triplet  excitation of the $SU(2)$ invariant Gross-Neveu model where the $S^z=0$ component is not given by a solution of the Bethe equations unlike the $S^z=\pm 1$ components.
 This state is  obtained by applying a spin lowering operator to the $S^z=1$ triplet excitation \cite{braak}. 
 In the present case due to the space parity symmetry we obtain a unique boundary string solution which occurs as a double pole in the Bethe equations of the STS phase, see Appendix  eq. (\ref{doublepole}). Hence one may expect that it should count as two states, namely, that there exists another state $|\widehat{0}'\rangle$ with $S^z=0$ in addition to the state $|\widehat{0}\rangle$. Such a state cannot be obtained simply by a lowering operator as was the case for the $SU(2)$ multiplets, discussed earlier, since $\mathbb{Z}_2$ representations are all one dimensional. 
 
 To circumvent this problem  we  break
 the space parity symmetry by considering asymmetric boundary conditions, which splits the double pole and yields another boundary string solution. As a result the second spin singlet state is obtained as a solution to the Bethe equations leading to two quasi-degenerate ground states in the limit of infinitesimal asymmetry
 with wave functions localized at either the left or the right edge. These two states,
 in  properly renormalized symmetric limit, account for  a twofold degeneracy of the ground state missed by the Bethe Ansatz analysis of the symmetric  case.

We consider now the following asymmetric OBC  
   \bea
\Psi_{Ra}(L/2)&=&-B_{ab}\Psi_{Lb}(L/2),\\ \Psi_{Ra}(-L/2)&=&-\Psi_{Lb}(-L/2),
\eea
where
\bea
\label{twist1}
B_{ab}=\frac{1}{\cosh(f/2)}\left(\begin{array}{cc} \cosh(f/2+i\epsilon)&0\\0&\cosh(f/2-i\epsilon)\end{array}\right),
\nonumber \\
\eea
and  $\epsilon > 0$ is an asymmetry parameter. The latter boundary conditions, which break both space parity and the $\mathbb{Z}_2$ symmetry (\ref{z2}), give back the symmetric OBC in the limit $\epsilon \rightarrow 0$. Remarkably enough the problem is still integrable when $\epsilon \neq 0$ and the resulting Bethe equations (see Appendix (\ref{sec:AppendixBE})) are given by
\bea\nonumber \small
-2\pi I_\alpha+\sum_{\sigma=\pm}N\Theta(\lambda_\alpha+\sigma f/2u ,1/2)-\Theta(\lambda_\alpha+i\tau\pi/2u,1/2)\\ \nonumber-\Theta(\lambda_\alpha+i\tau \pi/2u,(1-\epsilon^\prime)/2) \label{aniBAE}=\sum_{\beta=1}^{M} \sum_{\sigma=\pm}\Theta\left(\lambda_\alpha+\sigma \lambda_\beta,1\right),\\
\label{BEAASYM} 
 \eea
  where $\tau =(1,0)$ in the regions $A$ and $\widehat{A}$ respectively and
  \be
  \epsilon^\prime=2\epsilon/u.
  \label{epsilonprime}
  \ee
  As we can readily see, unlike \eqref{BEA}, the latter equations display    two different boundary terms. These equations can be solved by following the same procedure as for symmetric boundary conditions.
  In order to obtain a non trivial solution in the scaling limit, where $u \rightarrow 0$, one needs to take also simultaneously the  limit $\epsilon \rightarrow 0$ 
  with $\epsilon^\prime$ maintained fixed and small. In this limit it is $\epsilon^\prime$
  that plays the role of the physical asymmetry
  parameter.
  
  In the SSS phase,  we find that the ground state in the region $A$ and on the $GN$ line has total spin $S^z=0$ in the scaling limit and that there is no boundary string solutions just as in the symmetric case. In this phase the asymmetry in the boundary conditions plays a marginal role. This is not  the case in the STS phase.
  This is due to the fact that, in the symmetric case,
  the two degenerate ground states carry a non zero spin $S^z=\pm 1/2$. The asymmetry at  the right boundary, which distinguishes between
  up and down spins, will then  lift the degeneracy.
  Indeed for a positive $\epsilon > 0$ we find that the ground state has a spin $S^z=- 1/2$ while the state with $S^z=+ 1/2$ has a higher  energy.
  On top of that,  due to the presence of the two different boundary terms in (\ref{BEAASYM}), there exists now {\it two} different boundary strings at 
  \be
  \lambda=\pm i/2, \; 
  \lambda=\pm i(1-\epsilon^\prime)/2.
  \label{anistrings}
  \ee
  For a positive $\epsilon > 0$ one can only add these two boundary strings to
 the state with the higher spin $S^z=+ 1/2$. Doing that we end up with {\it two different} spin singlet states with $S^z=0$ 
 \bea
 |\widehat{0}\rangle_{\epsilon'}   &\rightarrow& \lambda=\pm i/2, \nonumber \\
 |\widehat{0}'\rangle_{\epsilon'} &\rightarrow& \lambda=\pm i(1-\epsilon^\prime)/2.
 \label{sz0epsilon}
 \eea
  The calculation of the energy of the two boundary strings (\ref{anistrings}), and hence of the two states (\ref{sz0epsilon}),  proceed as in the symmetric case. For both strings the above energy splits into a charge and a spin part: $E_B=E_{\text{charge}}+E_{\text{spin}}$. While the charge contribution is still 
 $E_{\text{charge}}=- \pi/2L$ for both strings (\ref{anistrings}),
 the spin contributions  are  different. In the limit of large system size it is zero (to order ${\cal O}(e^{-mL}))$ for the first string while
 the second string has a finite energy which is precisely 
 the energy splitting between the two $S^z=\pm 1/2$ states. Therefore, as far as the spin degrees of freedom are concerned, in the presence of a non zero $\epsilon > 0$, the ground state is twofold degenerate and consists of two states ($|\widehat{0}'\rangle_{\epsilon'}$, $|-\widehat{\frac{1}{2} }\rangle_{\epsilon'}$). The two other states  ($|\widehat{0}\rangle_{\epsilon'}$, $|\widehat{\frac{1}{2}}\rangle_{\epsilon'}$) have a higher energy $\delta E_{\epsilon^\prime}=m \sin(\epsilon^\prime \pi/2)$. The calculation of this energy and discussion about the structure of the ground states and associated symmetries in the presence of asymmetric boundary conditions goes beyond the scope of this work, hence it will be discussed in further works.

 When  $\epsilon' \rightarrow 0$, the energy splitting between these states $\delta E_{\epsilon^\prime}$ goes to zero,  and the two spin singlet states (\ref{sz0epsilon}) are quasi-degenerate in the limit of infinitesimally small asymmetry $\epsilon' \ll 1$. They correspond to two zero energy boundary bound state modes which are localized at the two ends of the system. Although the above analysis does not tells us about the status of the two states with respect to the $\mathbb{Z}_2$ (\ref{z2}), i.e. whether they are symmetric or antisymmetric under the action of $\tau$, it does tells us that there are {\it two} states in the spin singlet sector when $0 < \epsilon' \ll 1$. On physical grounds, we do not expect anything special to happen to {\it the number of states} in the symmetric limit which should be two when  $\epsilon' \rightarrow 0$. Of course, when $\epsilon'=0$, the two boundary strings (\ref{anistrings}) become identical and the two states in (\ref{sz0epsilon}) overlap. However, as in the XXZ
 spin chain \cite{skorik}, we expect that, in a suitable renormalized limit
 $\epsilon' \rightarrow 0$, the two states $|\widehat{0}\rangle_{\epsilon'}$ and $|\widehat{0}'\rangle_{\epsilon'}$ 
 yield  different spin singlet states $|\widehat{0}\rangle$ and $|\widehat{0}'\rangle$ in the symmetric limit.
 These two singlet states
 together with the two odd fermion parity $S^z=\pm 1/2$ spin states  account for the fourfold degeneracy found in the semi-classical analysis. However, as in the
 symmetric case, the present analysis cannot
explain the status of the two singlet states
 with respect to the $\mathbb{Z}_2$ symmetry
and hence we are unable to relate these states with the Majorana construction given in the preceding section. 
 We hope to come back to this non trivial issue in a further publication.

\section{Discussion}
\label{sec:discussion}

We have provided  the exact solution of the $U(1)$ Thirring model on a finite line segment with both symmetric  and asymmetric open boundary conditions (OBC). We showed that the fourfold ground state degeneracy found by semi-classical analysis  \cite{Keselman2015}
can be understood as being due to the presence of two zero energy boundary bound states  localized at the edges of the system. These bound states correspond to two  boundary strings solutions of the Bethe equations in the presence of slightly asymmetric OBC. Our results are consistent
with the semi-classical analysis   based on the presence  of  spin $\pm 1/4$ localized at the two edges of the system and support the fact that  the massless spin-triplet superconducting topological state, predicted in the  anisotropic regime $g_{\parallel} \gg 1$, survives strong quantum fluctuations at least
in the region $\widehat{A}$ and on the dual 
$\widehat{GN}$ line.

However, our Bethe Ansatz approach cannot track down the two zero energy  Majorana modes ($\gamma_{\cal L}, \gamma_{\cal R})$
(see Eqs.(\ref{lambdaeta})),  associated with the $\mathbb{Z}_2$ symmetry (\ref{z2}), which  are responsible for the topological order in a given fermion parity sector. Probing these Majorana modes would require a detailed  calculation of  the wave functions in real space associated with  the boundary bound states. This is a formidable task in the present fermionic field theory. However, related work on the XXZ spin chain \cite{skorik}, where the boundary bound states wave functions can be obtained  with asymmetric boundary fields,  suggests that one could possibly probe these Majorana modes in a suitable symmetric limit.

 Although the second spin singlet state $|\widehat{0'}\rangle$ was  obtained  by considering slightly asymmetric OBC, it is not a solution of the Bethe equation in the symmetric case. An alternative way would be to construct an analog of a lowering operator acting on a highest weight spin-1 state to obtain the $S^z$ member of the multiplet, though in itself it cannot be obtained as a solution of the Bethe Ansatz equation \cite{braak}. Similarly, in our present case one would need to construct a "raising" or "lowering" operator $\Gamma$ which, when acting on the singlet solution $|\widehat{0} \rangle$, gives the desired  state, i.e:
\be
\Gamma |\widehat{0} \rangle = |\widehat{0'} \rangle.
\ee
In the topological phase such an operator
would be provided by one of two zero energy  Majorana modes $\Gamma=(\gamma_{\cal L}, \gamma_{\cal R})$ of  Eqs.(\ref{lambdaeta})  associated with the $\mathbb{Z}_2$ symmetry (\ref{z2})
with $[\Gamma, H]=[\Gamma,  {\cal P}]=0$, 
$\{\Gamma, \tau \}=0$ and $\Gamma^2=1$.

At last but not least, it would be interesting
to understand what happens in the regions
C($\widehat{C}$). Although we have obtained
the Bethe equations in these regions we find that there are issues when one wants to obtain universal answers. This leaves open the question of the topological nature  of the STS phase in the region $\widehat{C}$.
We hope to come on all these topics in further works.

 \acknowledgements
 The work reported here was begun while N.A. was visiting the  IPhT  Saclay. He wishes to thank H. Saleur for his kind hospitality. 
P.A  thanks A. Kesselman and E. Berg for enlightening discussions. P.P acknowledges  illuminating discussions with C. Rylands.

\bibliography{refpaper}

\begin{widetext}

\appendix

 \section{Bethe Ansatz}
In this section we derive the Bethe equations of the model subject to the following asymmetric boundary conditions
\bea
\Psi_{Ra}(L/2)=-B_{ab}\Psi_{Lb}(L/2), \;\; \Psi_{Ra}(-L/2)=-\Psi_{Lb}(-L/2)
\eea

where \bea\label{twistap}
B_{ab}=\frac{1}{\cosh(f/2)}\left(\begin{array}{cc} \cosh(f/2+i\epsilon)&0\\0&\cosh(f/2-i\epsilon)\end{array}\right), \eea

and $\epsilon>0$ is an asymmetry parameter. The left boundary has the usual open boundary condition whereas the right boundary has a 'slightly twisted' boundary condition. This breaks the space parity and $\mathbb{Z}_2$ symmetry which gives rise to two fundamental boundary string solutions. Symmetric boundary condition can be obtained by taking the limit $\epsilon\rightarrow 0$ which restores the broken space parity and the $\mathbb{Z}_2$ symmetry.

\subsection{N-particle solution}
\label{sec:AppendixNparticles}

The Hamiltonian commutes with total particle number, $N=\int \psi_+^\dag(x)\psi_+(x)+\psi_-^\dag(x)\psi_-(x)$ and $H$ can be diagonalized by constructing the exact eigenstates in each $N$ sector. 
Since $N$ is a good quantum number we may construct the eigenstates by examining the different $N$ particle sectors separately. We start with $N=1$ wherein we can write the wavefunction as an expansion in plane waves,

\bea\nonumber
\ket{k}=\sum_{a_j=\uparrow\downarrow,\sigma=\pm}\int_{-\frac{L}{2}}^{\frac{L}{2}}\mathrm{d}x\, e^{i\sigma kx} A^\sigma_{a_1}    \psi^{\dagger}_{\sigma,a_1} (x)\ket{0} .
\eea
  $\ket{0}$ is the drained Fermi sea and $A^\sigma_{a_1}$ are the amplitudes for an electron with chirality $\sigma$ and spin $a_1$. The two boundary S-matrices $S^{1R}_{a_1b_1},S^{1L}_{a_1b_1}$ exchange the chirality of a particle. 
\bea A^-_{a_1}=S^{1R}_{a_1b_1} \; A^+_{b_1} \\A^+_{a_1}=S^{1L}_{a_1b_1} \; A^-_{b_1}.\eea

 The asymmetric boundary conditions \eqref{twistap} lead to the following boundary S-matrices \bea S^{1R}_{ab}=B^\dagger_{ab}, \;\; S^{1L}_{ab}=I_{ab}. \eea Applying the boundary condition at the left boundary also quantizes the bare particle momentum $k$.
 
 We now consider the two particle sector, $N=2$, were the bulk interaction plays a role.
Since the two particle interaction is point-like
we may divide configuration space into regions such that
the interactions only occur at the boundary between two
regions. Therefore away from these boundaries we write
the wave function as a sum over plane waves so that the most general two particle state can be written as
\bea\label{2particle}
\ket{k_1,k_2}&=& \sum_{\sigma,a} \int_{-\frac{L}{2}}^{\frac{L}{2}}\mathrm{d}^2x\,F_{a_1a_2}^{\sigma_1\sigma_2}(x_1,x_2)e^{\sum_{j=1}^2i\sigma_jk_jx_j}\psi^{\dagger}_{\sigma_1a_1}(x_1)\psi^{\dagger}_{\sigma_2a_2}(x_2) \ket{0},
\eea 
where we sum over all possible spin and chirality configurations  and the two particle wavefunction, $F_{a_1a_2}^{\sigma_1\sigma_2}(x_1,x_2)$ is split up according to the ordering of the particles,
\bea
 F_{a_1a_2}^{\sigma_1\sigma_2}=A_{a_1a_2}^{\sigma_1\sigma_2}[12]\theta(x_2-x_1)+A_{a_1a_2}^{\sigma_1\sigma_2}[21]\theta(x_1-x_2).
\eea
The amplitudes $A_{a_1a_2}^{\sigma_1\sigma_2}[Q]$ refer to a certain chirality and spin configuration, specified by $\sigma_j$, $a_j$ as well as an ordering of the particles in configuration space denoted by $Q$. For $Q=12$ particle $1$ is to the left of particle $2$ while for $Q=21$ the  order of the particles are exchanged.
Applying the Hamiltonian to \eqref{2particle} we find that it is an eigenstate with energy $E=k_1+k_2$ provided that these amplitudes are related to each other via application of $S$-matrices. The amplitudes which differ by exchanging the chirality of the leftmost or the rightmost particle are related by the boundary S-matrices.
\bea
A^{\sigma_1-}[12]=S^{2R} \;A^{\sigma_1+}[12], \;\; A^{+\sigma_2}[12]=S^{1L} \;A^{-\sigma_2}[12],\\
A^{-\sigma_2}[21]=S^{1R}\;A^{+\sigma_2}[21],\;\; A^{\sigma_1+}[21]=S^{2L}\;A^{\sigma_1-}[21].\eea
 As discussed above in the one particle case, the boundary S-matrices are  $S^{1R}=B^\dagger, \; S^{1L}=I,\; S^{2R}=B^\dagger, \; S^{2L}=I$. For ease of notation we have suppressed spin indices. It is understood that $S^{1R},S^{1L}$ act in the spin space of particle 1 whereas $S^{2R},S^{2L}$ act in the spin space of particle 2.

There are two types of two particle bulk $S$-matrices denoted by $S^{12}$ and $W^{12}$ which arise due to the bulk interactions and relate amplitudes which have different orderings. The first relates amplitudes which differ by exchanging the order of particles with opposite chirality 
\bea
A^{+-}[21]=S^{12}A^{+-}[12],\\
A^{-+}[12]=S^{12}A^{-+}[21],
\eea
where  $S^{12}$ acts on the spin spaces of particles 1 and 2. Explicitly it is given by, \cite{duty}
\bea\label{S12} S^{ij}= \left(\begin{array}{cccc} 1&&&\\&\frac{\sinh(f)}{\sinh(f+\eta)}&\frac{\sinh(\eta)}{\sinh(f+\eta)}&\\&\frac{\sinh(\eta)}{\sinh(f+\eta)}&\frac{\sinh(f)}{\sinh(f+\eta)}&\\&&&1\end{array}\right).\eea
where $\eta=-iu$ and $f$, $u$ are related to $g_{\parallel}$ and $g_{\perp}$ through the relations $\cos(u)=\frac{\cos(g_{\parallel})}{\cos(g_{\perp})}$ and $\frac{\sin(u)}{\tanh(f)}=\frac{\sin(g_{\parallel})}{\cos(g_{\perp})}.$ In obtaining the above form of the S matrix we have ignored an unimportant overall factor.
Whilst the second {{type of $S$-matrix}} relates amplitudes where  particles of the same chirality are exchanged,
\bea
A^{--}[21]=W^{12}A^{--}[12],\\\label{W12}
A^{++}[12]=W^{12}A^{++}[21].
\eea
 Unlike \eqref{S12}, $W^{12}$ is not fixed by the Hamiltonian but rather by the consistency of the construction. This is expressed through the Yang-Baxter equations 
\bea\label{BYB1}
S^{23}\;S^{13}\;W^{12}&=&W^{12}\;S^{13}\;S^{23},\\
W^{23} \;W^{13} \;W^{12}& = &W^{12} \;W^{13}\; W^{23},\\  S^{2R}\;S^{12}\;S^{1R}\;W^{12}&=&W^{12}\;S^{1R}\;S^{12}\;S^{2R},\\\label{BYB2}S^{2L}\;S^{12}\;S^{1L}\;W^{12}&=&W^{12}\;S^{1L}\;S^{12}\;S^{2L},\eea

which need to be satisfied for the eigenstate to be consistent. We take $W^{12}=P^{12}$ which can be explicitly checked to satisfy \eqref{BYB1}-\eqref{BYB2}. The relations \eqref{2particle}-\eqref{W12} provide a complete set of solutions of the two particle problem.

We can now generalize this to the $N$-particle sector and find that the eigenstates of energy $E=\sum_{j=1}^Nk_j$ are of the form
\bea\label{NparticleS}
\ket{\{k_j\}}=
\sum_{Q,\vec{a},\vec{\sigma}}\int \theta(x_Q) A^{\{\sigma\}}_{\{a\}}[Q] \prod_j^N e^{i\sigma_j k_jx_j}\psi^{\dagger}_{a_j\sigma_j}(x_j)\ket{0}.
\eea
Here we sum over all  spin and chirality configurations specified by $\{a\}=\{a_1\dots a_N\}$, $\{\sigma\}=\{\sigma_1\dots \sigma_N\}$ as well as different orderings of the $N$ particles. These different orderings correspond to elements of the symmetric group $Q\in \mathcal{S}_N$. In addition $\theta(x_Q)$ is the Heaviside function which is nonzero only for that particular ordering. 
As in the $N=2$ sector the amplitudes $A^{\vec{\sigma}}_{\vec{a}}[Q]$ are related to each other by the various $S$-matrices in the same manner as before i.e. amplitudes which differ by changing the chirality of the leftmost particle are equal as $S^{jL}=I$, the amplitudes which differ by changing the chirality of the rightmost particle are related by $S^{jR}$ and the amplitudes which differ by exchanging the order of opposite or same chirality particles are related by $S^{ij}$ and $W^{ij}$ respectively. The consistency of this construction is then guaranteed by virtue of these $S$-matrices satisfying the following Yang-Baxter equations\cite{Sklyannin, Cherednik, ZinnJustin}
\bea\label{YB1}
W^{jk} \;W^{ik}\; W^{ij} &=& W^{ij} \;W^{ik} \;W^{jk},\\ \label{YB2}
S^{jk}\;S^{ik}\;W^{ij} &=& W^{ij}\;S^{ik}\;S^{jk},
\\ S^{jR}\;S^{ij}\;S^{iR}\;W^{ij}&=&W^{ij}\;S^{iR}\;S^{ij}\;S^{jR}\label{YB3},\\S^{jL}\;S^{ij}\;S^{iL}\;W^{ij}&=&W^{ij}\;S^{iL}\;S^{ij}\;S^{jL}\label{YB4},\eea
 Where $W^{ij}=P^{ij}$ and as before the superscripts denote which particles the operators act upon.

\subsection{Bethe equations}
\label{sec:AppendixBE}

In this section we derive the Bethe equations (3). 
Enforcing the boundary condition at $x=-L/2$ on the eigenstate \eqref{NparticleS} we obtain the following eigenvalue problem which constrains the $k_j$,
\bea
e^{-2ik_jL}A^{\{\sigma\}}_{\{a\}}[\mathbb{1}]=\left(Z_j\right)^{\{\sigma\},\{\sigma\}'}_{\{a\},\{a\}'} A^{\{\vec{\sigma}'\}}_{\{\vec{a}'\}}[\mathbb{1}].
\eea
Here $\mathbb{1}$ denotes the identity element of $\mathcal{S}_N$, i.e. $\mathbb{1}=12\dots N$ and the operator $Z_j$ is the transfer matrix for the $j^\text{th}$ particle given by
\bea
Z^j=W^{jj-1}\dots W^{j1} S^{j1}...S^{jj-1}S^{jj+1}...S^{jN}S^{jR}W^{jN}...W^{jj+1}\eea
where the spin indices have been suppressed. This operator takes the $j^\text{th}$ particle from one side of the system to the other and back again, picking up $S$-matrix factors along the way as it moves past the other $N-1$ particles, first as a right mover and then as a left mover.   Using the relations \eqref{YB1}- \eqref{YB4} one can prove that all the transfer matrices commute, $[Z^j,Z^k]=0$ and therefore are simultaneously diagonalizable. In order to determine the spectrum of $H$ we must therefore diagonalize $Z^j,~\forall j$. Here we choose to diagonalize $Z^1$. To do this we use the method of boundary algebraic Bethe Ansatz \cite{Sklyannin, Cherednik, ODBA}.  In order to use this method we need to embed the bare S-matrices in a continuum \cite{trieste} that is, we need to find the matrices $R(\lambda)$, $K(\lambda)$ such that for certain values of the spectral parameter $\lambda$, we obtain the bare S-matrices of our model. Note that the S matrix $S^{12}$ is of the form of $XXZ$ $R$ matrix 

\bea R^{ij}(\lambda)=\left(\begin{array}{cccc} 1&&&\\&\frac{\sinh(\lambda)}{\sinh(\lambda+\eta)}&\frac{\sinh(\eta)}{\sinh(\lambda+\eta)}&\\&\frac{\sinh(\eta)}{\sinh(\lambda+\eta)}&\frac{\sinh(\lambda)}{\sinh(\lambda+\eta)}&\\&&&1\end{array}\right).\eea

 We can see that \small{$R^{ij}(0)=W^{ij}, \hspace{2mm} R^{ij}(f)= S^{ij}$}. The $K$ matrix is given by \cite{Sklyannin} \bea K^j(\lambda)= \frac{1}{\cosh(\lambda)}\left(\begin{array}{cc} \cosh(\lambda-i\epsilon)&0\\0&\cosh(\lambda+i\epsilon)\end{array}\right)\eea and it is related to the right boundary S-matrix as $S^{jR}=K^j(f/2)$.
 The transfer matrix{{ $Z_1$ }}is related to the Monodromy matrix $\Xi_{\tau}(\lambda)$ as $Z^1=t(\frac{f}{2})=\Tr_{\tau} \Xi^{\tau}(\frac{f}{2})$, where
\bea\Xi^{\tau}(\lambda)= R^{1\tau}(\lambda+\frac{f}{2})...R^{N\tau}(\lambda+\frac{f}{2})K^{\tau}(\lambda)R^{N\tau}(\lambda-\frac{f}{2})...R^{1\tau}(\lambda-\frac{f}{2}).\eea
Here $\tau$ represents {{an  }}auxiliary space and $\Tr_\tau$ represents the trace in the auxiliary space.
{{Using the properties of the $R$ matrices one can prove that $[t(\lambda),t(\mu)]=0$ \cite{ODBA} and by expanding $t(\mu)$ in powers of $\mu$, obtain infinite set of conserved charges which guarantees integrability}}. By following the Boundary Algebraic Bethe Ansatz approach we obtain the following Bethe equations in the region $A$, corresponding to the reference state with all up spins 

\bea\label{energy}
e^{2ik_jL}=\beta^{-1}(f/2)\;\Pi_{\alpha=1}^M\Pi_{\sigma=\pm} \;  \gamma(f/2,\sigma\lambda_\alpha,u/2), \;\; \\ \gamma(x,y,z)=\frac{\sinh(x+y-iz)}{\sinh(x+y+iz)}, \beta(x)=\frac{\cosh(x-i\epsilon)}{\cosh(x)}
\eea

 where $\lambda_\alpha$, $\alpha=1,\dots,M$ are the Bethe roots which satisfy the following equations 
\bea\label{BAE}
\Pi_{\sigma=\pm}\gamma^N(\lambda_\alpha,\sigma f/2 ,u/2)\gamma(\lambda_\alpha,i\pi/2,-u/2)\gamma(\lambda_\alpha,i\pi/2,-(u-2\epsilon)/2)=\Pi_{\beta=1,\sigma=\pm}^{M} \gamma(\lambda_\alpha,\sigma \lambda_\beta,u).
\eea
 
 By rescaling $\lambda_\alpha\rightarrow u\lambda_\alpha$ and applying logarithm we obtain the following Bethe equations in the region $A$ with asymmetric boundary conditions.
 
 \bea\label{aniBE1}
 \sum_{\sigma=\pm}N\Theta(\lambda_\alpha+\sigma f/2u ,1/2)-\Theta(\lambda_\alpha+i\pi/2u,1/2) -\Theta(\lambda_\alpha+i\pi/2u,(1-\epsilon^\prime)/2)=\sum_{\beta=1}^{M} \sum_{\sigma=\pm}\Theta\left(\lambda_\alpha+\sigma \lambda_\beta,1\right)+2i\pi I_\alpha
 \eea

\bea
 k_j=\frac{\pi n_j}{L}+\frac{i}{2L}\left(\log[\beta(f/2)]+\sum_{\beta=1}^M\sum_{\sigma=\pm}\Theta(f/2u+\sigma\lambda_\beta,1/2)\right),
\eea

where $\displaystyle{\Theta(x,y)=\log\left(\frac{\sinh(u(x+iy))}{\sinh(u(x-iy))}\right)}$ and $\epsilon^\prime=2\epsilon/u$. To obtain the Bethe equations in the topological region $\widehat{A}$, we can work with $g_\parallel,g_\perp<0$ in the Hamiltonian and then take the limit $f\rightarrow -i\pi-f$ \cite{Japaridze} in the obtained Bethe equations. We obtain a different set of Bethe equations

\bea\label{anient}
e^{2ik_jL}=\widehat{\beta}^{-1}(f/2)\;\Pi_{\alpha=1}^M\Pi_{\sigma=\pm} \;  \gamma(f/2,\sigma\lambda_\alpha,u/2), \;\;   \widehat{\beta}(x)=\frac{\sinh(x-i\epsilon)}{\sinh(x)}
\eea

 \bea\label{aniBAET}
\Pi_{\sigma=\pm}\gamma^N(\lambda_\alpha,\sigma f/2 ,u/2) \gamma(\lambda_\alpha,0,-u/2)\gamma(\lambda_\alpha,0,-(u-2\epsilon)/2)=\Pi_{\beta=1,\sigma=\pm}^{M} \gamma(\lambda_\alpha,\sigma \lambda_\beta,u),
\eea
 Applying logarithm to the above equation and rescaling the Bethe roots we obtain the Bethe equations in the region $\widehat{A}$ with asymmetric boundary conditions,
\bea\label{aniBE2}
 \sum_{\sigma=\pm}N\Theta(\lambda_\alpha+\sigma f/2u ,1/2)- \Theta(\lambda_\alpha,1/2)-\Theta(\lambda_\alpha,(1-\epsilon^\prime)/2) =\sum_{\beta=1}^{M} \sum_{\sigma=\pm}\Theta\left(\lambda_\alpha+\sigma \lambda_\beta,1\right)+2i\pi I_\alpha,
 \eea

\bea
 k_j=\frac{\pi n_j}{L}+\frac{i}{2L}\left(\log[\widehat{\beta}(f/2)]+\sum_{\beta=1}^M\sum_{\sigma=\pm}\Theta(f/2u+\sigma\lambda_\beta,1/2)\right).
\eea

The Bethe equations corresponding to the reference state with all down spins can be obtained by taking the limit  $\epsilon \rightarrow -\epsilon^\prime$ \cite{Sklyannin} in the above Bethe equations. Note that when symmetric boundary conditions are applied the Bethe equations corresponding to the reference state with all up spins are same as those corresponding to the reference state with all down spins. 

As already mentioned in the maintext the asymmetric boundary conditions \eqref{twistap} break the $\mathbb{Z}_2$ symmetry, this shifts the spin $S^z$ of all the states in the regions $A$ and $\widehat{A}$ by a term which is proportional to $\epsilon$. In the scaling limit $u \ll 1$, one also needs to take the limit $\epsilon \ll 1$ while holding $\epsilon^\prime=2\epsilon/u$ fixed. In this limit the shift in the values of $S^z$ of all the states goes to zero, and therefore the ground state in any certain region of the phase diagram with asymmetric boundary condition has the same total spin $S^z$ as that in the corresponding region with symmetric boundary condition. 

The profound effect of applying the asymmetric boundary conditions is that we now have two fundamental boundary string solutions $\lambda=\pm i/2,\;  \lambda=\pm i/2(1-\epsilon^\prime)$. 

To obtain the Bethe equations with symmetric boundary conditions in region $A$ we can take the limit $\epsilon^\prime\rightarrow 0$ in the equations \eqref{BAE}. We get
\bea
e^{2ik_jL}=\;\Pi_{\alpha=1}^M\Pi_{\sigma=\pm} \;  \gamma(f/2,\sigma\lambda_\alpha,u/2), \eea

\bea
\Pi_{\sigma=\pm}\gamma^N(\lambda_\alpha,\sigma f/2 ,u/2)\gamma^2(\lambda_\alpha,i\pi/2,-u/2)=\Pi_{\beta=1,\sigma=\pm}^{M} \gamma(\lambda_\alpha,\sigma \lambda_\beta,u).\eea

To obtain the Bethe equations with symmetric boundary conditions in region $\widehat{A}$ we can take the limit $\epsilon^\prime\rightarrow 0$ in the equations \eqref{aniBAET}. We get
\bea
e^{2ik_jL}=\;\Pi_{\alpha=1}^M\Pi_{\sigma=\pm} \;  \gamma(f/2,\sigma\lambda_\alpha,u/2), \eea

\bea \label{doublepole}
\Pi_{\sigma=\pm}\gamma^N(\lambda_\alpha,\sigma f/2 ,u/2) \gamma^2(\lambda_\alpha,0,-u/2)=\Pi_{\beta=1,\sigma=\pm}^{M} \gamma(\lambda_\alpha,\sigma \lambda_\beta,u).
\eea

\end{widetext}
\end{document}